\documentclass[lettersize,journal]{IEEEtran}
\usepackage{cite}
\usepackage{amsfonts,amsmath,color,amssymb,amsxtra,amsbsy,floatflt}
\usepackage{algorithmic}
\usepackage{graphicx}
\usepackage{textcomp}
\usepackage{xcolor}
\def\BibTeX{{\rm B\kern-.05em{\sc i\kern-.025em b}\kern-.08em
    T\kern-.1667em\lower.7ex\hbox{E}\kern-.125emX}}

\usepackage{enumitem}
\newlist{myenum}{enumerate}{3}
\setlist[myenum,1]{label*=\arabic*)}
\setlist[myenum,2]{label=\arabic{myenumi}.\arabic*)}
\setlist[myenum,3]{label=\arabic{myenumi}.\arabic{myenumii}.\arabic*)}

\def\1{\mathbf{1}}

\usepackage[acronym]{glossaries}
\usepackage{subfigure}
\usepackage[ruled,vlined,linesnumbered]{algorithm2e}
\usepackage{placeins}
\usepackage{epstopdf}
\usepackage{enumitem}

\usepackage{tikz}
\usetikzlibrary{automata, arrows, positioning,calc}

\usepackage{arydshln}

\usepackage{multirow}
\usepackage[export]{adjustbox}
\usepackage{xfrac}
\usepackage[hidelinks]{hyperref}

\newcommand{\name}{MAREA}

\usepackage{fancyhdr}
\usepackage{tcolorbox}

\begin{document}

\thispagestyle{empty} 

\vfill 

\begin{center}
\begin{tcolorbox}[
    colback=gray!10, colframe=black, fonttitle=\bfseries,
    width=0.9\textwidth, boxrule=1pt, arc=5pt, outer arc=5pt,
    boxsep=10pt, left=10pt, right=10pt, top=10pt, bottom=10pt
]
\centering
\textbf{THIS IS AN AUTHOR-CREATED POSTPRINT VERSION.}

\vspace{0.3cm}

\textbf{Disclaimer:}  
This work has been accepted for publication in \textit{IEEE Transactions on Communications}.  

\vspace{0.3cm}

\textbf{Copyright:}  
© 2025 IEEE. Personal use of this material is permitted. Permission from IEEE must be obtained for all other  
uses, in any current or future media, including reprinting/republishing this material for advertising or  
promotional purposes, creating new collective works, for resale or redistribution to servers or lists, or reuse of  
any copyrighted component of this work in other works.

\vspace{0.3cm}

\textbf{DOI:} \href{https://doi.org/10.1109/TCOMM.2025.3552296}{10.1109/TCOMM.2025.3552296}

\end{tcolorbox}
\end{center}

\vfill 

\clearpage
\setcounter{page}{1}


\newacronym{3GPP}{3GPP}{3rd Generation Partnership Project}

\newacronym{5G}{5G}{5th Generation}
\newacronym{5G-ACIA}{5G-ACIA}{5G Alliance for Connected Industries and Automation}
\newacronym{5G-NR}{5G-NR}{5G New Radio}
\newacronym{6G}{6G}{Sixth Generation}

\newacronym{AMC}{AMC}{Adaptive Modulation and Coding}
\newacronym{AC}{AC}{Admission Control}
\newacronym{AGV}{AGV}{Automated Guided Vehicle}
\newacronym{AR}{AR}{Augmented Reality}

\newacronym{BLER}{BLER}{Block Error Rate}
\newacronym{BWP}{BWP}{Bandwidth Part}
\newacronym{BS}{BS}{Base Station}
\newacronym{BSS}{BSS}{Business Support System}

\newacronym{CDF}{CDF}{Cumulative Distribution Function}
\newacronym{CCDF}{CCDF}{Complementary Cumulative Distribution Function}
\newacronym{CDMA}{CDMA}{Code Division Multiple Access}
\newacronym{CTMC}{CTMC}{Continuos-Time Markov Chain}
\newacronym{CSI}{CSI}{Channel State Information}
\newacronym{CP}{CP}{Control Plane}
\newacronym{CQI}{CQI}{Channel Quality Indicator}
\newacronym{CU}{CU}{Centralized Unit}

\newacronym{DL}{DL}{downlink}
\newacronym{DNC}{DNC}{Deterministic Network Calculus}
\newacronym{DRB}{DRB}{Dedicated Radio Bearer}
\newacronym{DRP}{DRP}{Dynamic Resource Provisioning}
\newacronym{DRL}{DRL}{Deep Reinforcement Learnning}
\newacronym{DU}{DU}{Distributed Unit}

\newacronym{eMBB}{eMBB}{enhanced Mobile Broadband}
\newacronym{ETSI}{ETSI}{European Telecommunication Standards Institute}
\newacronym{EBB}{EBB}{Exponentially Bounded Burstiness}
\newacronym{EBF}{EBF}{Exponentially Bounded Fluctuation}
\newacronym{E2E}{E2E}{End-to-End}
\newacronym{EDF}{EDF}{Earliest Deadline First}
\newacronym{EM}{EM}{Expectation-Maximization}

\newacronym{FCAPS}{FCAPS}{Fault, Configuration, Accounting, Performance and Security}
\newacronym{FCFS}{FCFS}{First-come First-served}
\newacronym{FIFO}{FIFO}{First In First Out}

\newacronym{GBR}{GBR}{Guaranteed Bit Rate}
\newacronym{GMM}{GMM}{Gaussian Mixture Model}
\newacronym{GSMA}{GSMA}{Global System for Mobile Communications Association}
\newacronym{GST}{GST}{Generic Network Slice Template}
\newacronym{gNB}{gNB}{Next generation NodeB}

\newacronym{HDR}{HDR}{High Data Rate}

\newacronym{ITU}{ITU}{International Telecommunication Union}
\newacronym{IoT}{IoT}{Internet of Things}
\newacronym{ILP}{ILP}{Integer Linear Programming}
\newacronym{ICIC}{ICIC}{Inter-Cell Interference Cancellation}

\newacronym{LA}{LA}{Link Adaptation}
\newacronym{LOS}{LoS}{Line-of-Sight}
\newacronym{LSTM}{LSTM}{Long Short-Term Memory}
\newacronym{LTE}{LTE}{Long Term Evolution}

\newacronym{MAC}{MAC}{Medium Access Control}
\newacronym{MEC}{MEC}{Multi-access Edge Computing}
\newacronym{MCS}{MCS}{Modulation and Coding Scheme}
\newacronym{MDN}{MDN}{Mixture Density Network}
\newacronym{MGF}{MGF}{Moment Generating Function}
\newacronym{MIMO}{MIMO}{Multiple Input Multiple Output}
\newacronym{MISO}{MISO}{Multiple Input Single Output}
\newacronym{ML}{ML}{Machine Learning}
\newacronym{MNO}{MNO}{Mobile Network Operator}
\newacronym{mMTC}{mMTC}{Machine Type Communication}
\newacronym{MSE}{MSE}{Mean Squared Error}
\newacronym{mURLLC}{mURLLC}{massive ultra-Reliable Low Latency Communication}

\newacronym{NE}{NE}{Nash Equilibrium}
\newacronym{NEST}{NEST}{Network Slice Type}
\newacronym{NIP}{NIP}{Non-linear Integer Programming}
\newacronym{NFMF}{NFMF}{Network Function Management Function}
\newacronym{NFV}{NFV}{Network Function Virtualization}
\newacronym{NG-RAN}{NG-RAN}{Next Generation - RAN}
\newacronym{NLOS}{NLoS}{Non-Line-of-Sight}
\newacronym{NN}{NN}{Neural Network}
\newacronym{NSO}{NSO}{Network Slice Orchestrator}
\newacronym{NSMF}{NSMF}{Network Slice Management Function}
\newacronym{NSSMF}{NSSMF}{Network Slice Subnet Management Function}
\newacronym{NR}{NR}{New Radio}

\newacronym{OFDMA}{OFDMA}{Orthogonal Frequency-Division Multiple Access}
\newacronym{O-RAN}{O-RAN}{Open Radio Access Network}
\newacronym{PDF}{PDF}{Probability Density Function}
\newacronym{PMF}{PMF}{Probability Mass Function}
\newacronym{PRB}{PRB}{Physical Resource Block}
\newacronym{P-NEST}{P-NEST}{private NEST}

\newacronym{QoS}{QoS}{Quality of Service}

\newacronym{RAN}{RAN}{Radio Access Network}
\newacronym{RB}{RB}{Resource Block}
\newacronym{RBG}{RBG}{Resource Block Group}
\newacronym{RIC}{RIC}{RAN Intelligent Controller}
\newacronym{RIS}{RIS}{Reconfigurable Intelligent Surfaces}
\newacronym{RLC}{RLC}{Radio Link Control}
\newacronym{RRM}{RRM}{Radio Resource Management}
\newacronym{RRC}{RRC}{Radio Resource Control}
\newacronym{RSRP}{RSRP}{Received Signal Received Power}
\newacronym{RSRQ}{RSRQ}{Received Signal Received Quality}
\newacronym{RSSI}{RSSI}{Received Signal Strength Indication}
\newacronym{RT}{RT}{Real Time}
\newacronym{RU}{RU}{Radio Unit}

\newacronym{SDO}{SDO}{Standards Developing Organization}
\newacronym{SINR}{SINR}{Signal-to-Interference-plus-Noise Ratio}
\newacronym{SLA}{SLA}{Service Level Agreement}
\newacronym{SMO}{SMO}{Service Management and Orchestration}
\newacronym{SNC}{SNC}{Stochastic Network Calculus}
\newacronym{SNR}{SNR}{Signal-to-Noise Ratio}
\newacronym{S-NEST}{S-NEST}{standardized NEST}

\newacronym{TTI}{TTI}{Transmission Time Interval}

\newacronym{UE}{UE}{User Equipment}
\newacronym{UL}{UL}{Uplink}
\newacronym{UP}{UP}{User Plane}
\newacronym{uRLLC}{uRLLC}{ultra-Reliable Low Latency Communication}

\newacronym{V2X}{V2X}{Vehicle-to-Everything}
\newacronym{VBR}{VBR}{Variable Bit Rate}
\newacronym{VR}{VR}{Virtual Reality}
\newacronym{vRAN}{vRAN}{virtualized RAN}
\newacronym{vBS}{vBS}{virtualized Base Station}
\newacronym{VS}{VS}{Validation Scenario}

\newacronym{WiMAX}{WiMAX}{Worldwide Interoperability for Microwave Access}
\newacronym{WCDMA}{WCDMA}{Wideband \gls{CDMA}}

\title{\name{}: A Delay-Aware Multi-time-Scale Radio Resource Orchestrator for 6G O-RAN

\thanks{This work is part of grant PID2022-137329OB-C43 funded by MICIU/AEI/ 10.13039/501100011033. It has also been financially supported by the Ministry for Digital Transformation and of Civil Service of the Spanish Government through TSI-063000-2021-28 (6G-CHRONOS) project, and by the European Union through the Recovery, Transformation and Resilience Plan - NextGenerationEU, and in part by SNS JU Project 6G-GOALS (GA no. 101139232).

Oscar Adamuz-Hinojosa is with the Department of Signal Theory, Telematics and Communications, University of Granada, Granada, Spain (e-mail: oadamuz@ugr.es). Lanfranco Zanzi, Vincenzo Sciancalepore, and Xavier Costa-Perez are with NEC Laboratories Europe, Heidelberg, Germany. (e-mail: \{name.surname\}@neclab.eu). Xavier Costa-Perez is also with i2CAT Foundation and ICREA, Barcelona, Spain (e-mail:\{name.surname\}@i2cat.net).}

}

\author{\IEEEauthorblockN{
Oscar Adamuz-Hinojosa,
Lanfranco Zanzi,
Vincenzo Sciancalepore,
Xavier~Costa-Pérez
}

}



\maketitle

\begin{abstract}
The Open Radio Access Network (O-RAN)-compliant solutions often lack crucial details for implementing effective control loops at various time scales. 
To overcome this, we introduce \name{}, an O-RAN-compliant mathematical framework designed for the allocation of radio resources to multiple ultra-Reliable Low Latency Communication (uRLLC) services. 
In the near-real-time (RT) control loop, \name{} employs a novel Martingales-based model to determine the guaranteed radio resources for each uRLLC service. Unlike traditional queueing theory approaches, this model ensures that the probability of packet transmission delays exceeding a predefined threshold---the violation probability---remains below a target tolerance. 

Additionally, \name{} uses a real-time control loop to monitor transmission queues and dynamically adjust guaranteed radio resources in response to traffic anomalies. To the best of our knowledge, \name{} is the first O-RAN-compliant solution that leverages Martingales for both near-RT and RT control loops. Simulations demonstrate that \name{} significantly outperforms reference solutions, achieving an average violation probability that is $\times 10$ lower.
\end{abstract}


\section{Introduction}
In \gls{6G} networks, a critical challenge is managing the coexistence of multiple \gls{uRLLC} services, which impose stringent latency and reliability requirements to ensure seamless and efficient operations~\cite{Popovski2018}. One of the primary advancements driving the evolution of 6G is the virtualization of the \gls{RAN}~\cite{Tang2023}, achieved through the deployment of \gls{vRAN} instances. These instances consist of fully configurable \glspl{vBS}, tailored to the specific demands of diverse communication services.

{\bf Background.}
To address the increasing complexity of these networks, the \gls{O-RAN} Alliance has introduced an innovative, flexible architecture~\cite{Abdalla2022} that integrates the \gls{3GPP} functional split. This architecture distributes the \gls{vBS} across multiple network nodes, including the \gls{CU}-\gls{CP}, \gls{CU}-\gls{UP}, \gls{DU} and \gls{RU}. 
Additionally, \gls{O-RAN} enhances network agility and autonomy through its unique inclusion of two \glspl{RIC}: the non-\gls{RT} \gls{RIC}, which handles long-term network optimizations like policy computation and \gls{ML} model management via third-party applications (\emph{rApps}), and the near-\gls{RT} \gls{RIC}, which focuses on near-real-time RAN optimization and control (from $10$ ms to $1$ second) through \emph{xApps}.

\gls{O-RAN} also introduces three critical interfaces: the \texttt{A1} interface, enabling non-\gls{RT} \gls{RIC} to manage policies and ML models for long-term optimization; the \texttt{O1} interface, which allows the \gls{SMO} framework to oversee \gls{FCAPS} tasks for the \gls{vBS} and near-\gls{RT}-\gls{RIC}; and the \texttt{E2} interface, designed for near-\gls{RT}-\gls{RIC} to gather performance metrics from \gls{vBS} components and make rapid adjustments to resource allocation and service quality. For more details on \gls{O-RAN}, refer to ~\cite{survey-O-RAN}.

However, despite these advances, several challenges persist in fully realizing the potential of \gls{O-RAN}, particularly in the context of uRLLC services. The inherent architectural design of \gls{O-RAN} imposes limitations on making ultra-low latency decisions, such as packet scheduling for \gls{uRLLC}, which demands sub-millisecond response times~\cite{LACO}. Moreover, limited access to fine-grained, low-level information via the \texttt{E2} interface (e.g., transmission queue states, real-time channel quality) and the communication latencies involved in gathering this information further compound the difficulty of achieving real-time optimization within the required timescales~\cite{dApps_article}.

{\bf Motivation.}
This calls for an innovative \gls{RT} control loop to monitor and orchestrate \gls{MAC} schedulers, especially in scenarios where multiple \glspl{vBS} are deployed. In such cases, each \gls{vBS} may operate with a dedicated \gls{DU} featuring customized network functionalities. Building on the ongoing study of \gls{RT} control loop mechanisms~\cite{dApps_article}, \emph{this paper establishes a foundation} for future research into a comprehensive \gls{RT} orchestration framework.

One of the key challenges in \gls{O-RAN}-compliant networks is ensuring that each \gls{uRLLC} service meets its stringent delay requirement throughout its lifecycle. \textit{This can be seen as an optimization problem, where the objective is to minimize for each service the probability the packet transmission delay $w$ exceeds a given bound $W$}, i.e., $\min\left\{ \mathbb{P}[w>W]\right\}$, \textit{while ensuring that it remains below a predefined threshold} $\varepsilon$, i.e., $\mathbb{P}[w>W] \leq \varepsilon$. To achieve this, it is necessary to determine the optimal allocation of radio resources for each \gls{uRLLC} service at the near-\gls{RT} scale. Additionally, the \gls{RT} control loop must dynamically adjust the radio resource allocation computed in near-\gls{RT} scale to prevent violations of these delay constraints due to unexpected traffic variations. Therefore, the effective coordination of near-\gls{RT} and \gls{RT} control loops is essential for optimizing radio resource allocation and maintaining the required performance for coexisting \gls{uRLLC} services~\cite{survey-O-RAN}.
\color{black}


To ensure a packet's transmission delay \textcolor{black}{$w$} stays within a bound $W$ with a violation probability $\varepsilon$, i.e., $\mathbb{P}[w>W]<\varepsilon$ it is crucial to estimate \textcolor{black}{$\mathbb{P}[w>W]$} accurately. \gls{SNC} is a powerful tool that allows to calculate this probability under various traffic and cell capacity conditions. Previous work~\cite{SOTANC1,SOTANC2,SOTANC3} applied \gls{SNC} to estimate delay bounds for specific radio resource allocations in individual \gls{uRLLC} services. In~\cite{Adamuz-Hinojosa-TWC2023}, we extended \gls{SNC} for multi-service resource planning, though this approach, which dedicated \glspl{RB} per service, risked inefficient resource overprovisioning.
Our latest work introduced an O-RAN-compliant framework~\cite{Adamuz2024} that builds on \gls{SNC}, optimizing radio resource allocation across multiple \gls{uRLLC} services using historical traffic and capacity data. Unlike earlier methods, we integrate \gls{O-RAN} control loops, including a near-\gls{RT} \gls{SNC}-based controller that dynamically determines guaranteed \glspl{RB} for each service to maintain violation probabilities within target limits. Additionally, we propose a \gls{RT} control loop that adjusts \glspl{RB} in response to transmission queue data, addressing traffic anomalies and further reducing delay violation probabilities while improving resource efficiency. For further details, we refer the reader to Section~\ref{sec:RelatedWorks}.

{\bf Contributions.} While \gls{SNC}-based models provide clear advantages in terms of latency guarantee, their conservative delay bound estimations may limit the number of deployable services~\cite{Fidler1}.
To address this point, in this work, we explore the Martingales queueing methodology, which has shown notable improvements in delay bound estimation~\cite{Ciucu2014,Poloczek2015}. \textcolor{black}{Specifically, we propose a framework that builds upon our previous research~\cite{Adamuz2024}, where we used an \gls{SNC}-based approach, and extend it by incorporating a model based on Martingales Theory.} Several research works, such as~\cite{Zhao2018,Yan2024,Peng2024,Yu2024}, have proposed radio resource allocation solutions using Martingales-Theory-based models for estimating delay bounds. However, many of these solutions depend on well-known statistical distributions, e.g., Poisson, Bernoulli, Markov on-off, which hardly match real traffic patterns that may be the realization of arbitrary distributions.
Conversely, this work sheds the light on adapting a Martingales-Theory-based model to arbitrary traffic conditions. Specifically, we focus on the \gls{DL} operation of a single cell supporting multiple \gls{uRLLC} services, each with specific packet delay budget and violation probability requirements. Nevertheless, our solution is adaptable to broader scenarios involving \gls{UL} transmissions and multiple cells. The main contributions can be summarized as follows:
\begin{itemize}
    \item[(C1)] Building on the framework and findings of ~\cite{Adamuz2024}, we introduce \name{} as a complementary approach to support near-\gls{RT} and \gls{RT} control loops that can adapt to arbitrary traffic and channel statistical distributions. \name{} incorporates a novel Martingales-based controller, replacing the original \gls{SNC}-based controller in the near-\gls{RT} control loop. This enhancement improves the allocation of guaranteed \glspl{RB} for each \gls{uRLLC} service while maintaining the probability of delay violation below a given target. 
    \item[(C2)] We provide a detailed description of the functional and building blocks that constitute the \name{} framework, focusing on its integration with the \gls{O-RAN} architecture and operations in both near-\gls{RT} and \gls{RT} control loops. We also provide valuable insights related to \name{}'s capabilities and interaction with the \gls{O-RAN} ecosystem.
    \item[(C3)] We provide a comprehensive evaluation of \name{} by means of an exhaustive simulation campaign involving realistic 5G-\gls{NR} scenarios, comparing the Martingales-based approach against the previous proposed \gls{SNC}-based model. Our results show that \name{} effectively achieves more accurate delay-bound estimations, which translates into the possibility to safely support a larger number of uRLLC services for the given cell configuration. Our evaluation also demonstrates improved execution times, making \name{} better suited for real-world applications.
\end{itemize}


The remainder of this paper is organized as follows. Section~\ref{sec:OurFramework} defines \name{}. Section~\ref{sec:MartingaleTheory} introduces the proposed Martingales-based model. In Section~\ref{sec:SNC-basedOrchestrator}, we describe how \name{} executes the multi-time-scale control loops. Section~\ref{sec:PerformanceResults} evaluates \name{}'s performance. \textcolor{black}{Section~\ref{sec:Challenges} discuss key challenges for implementing \name{} in commercial O-RAN deployments}. Section~\ref{sec:RelatedWorks} discusses related works. Finally, Section~\ref{sec:Conclusions} concludes the paper.

\section{The \name{} Framework}\label{sec:OurFramework}
\begin{figure}[t!]
    \centering
    \includegraphics[width=\columnwidth]{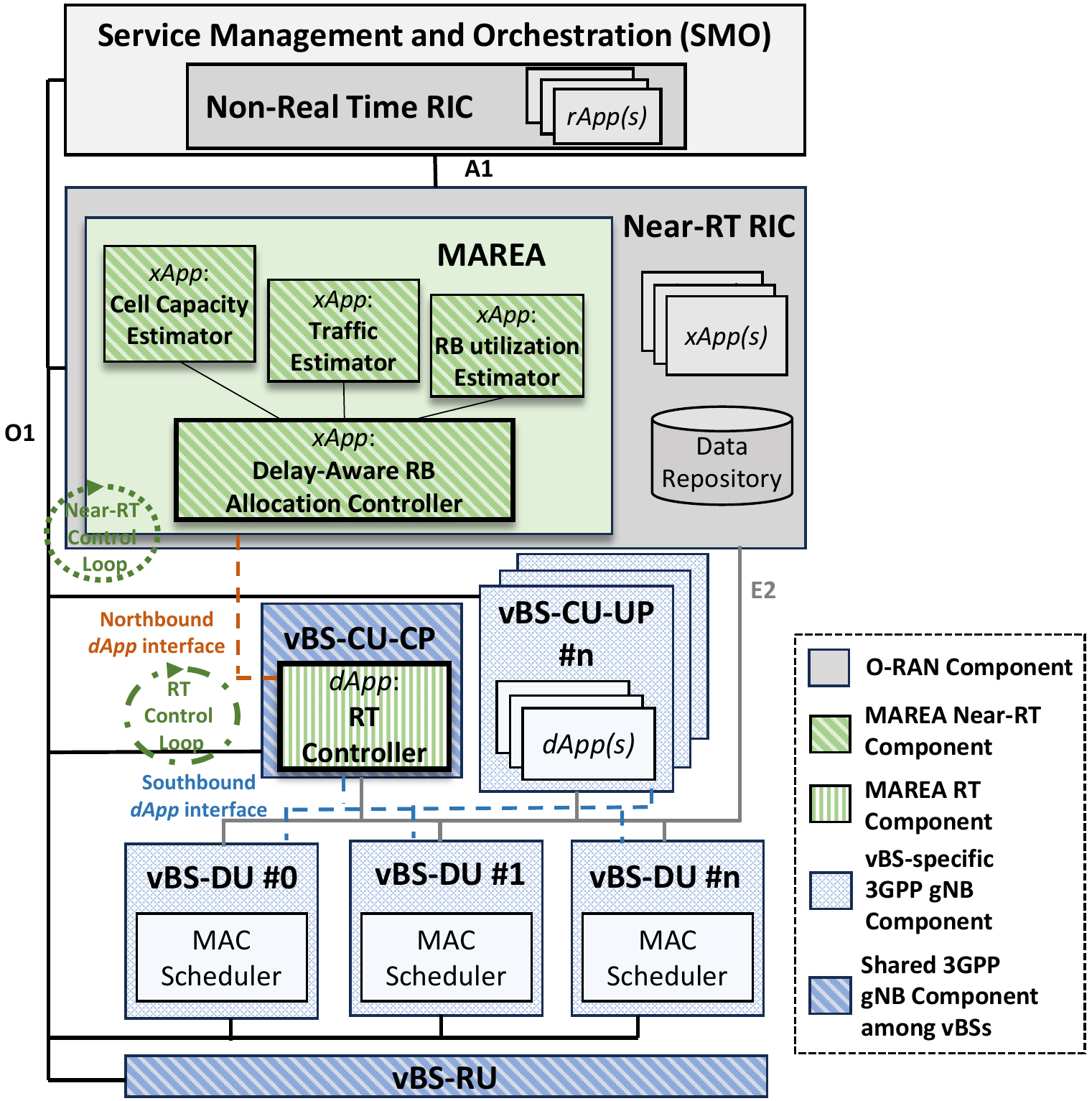}
    \caption{Integration of \name{} (i.e., green blocks) in the O-RAN architecture. For simplicity, only the northbound and southbound interfaces of the \gls{RT} Controller \emph{dApp} are shown. Other \emph{dApps} would have similar interfaces.}
    \label{fig:FrameworkIntegration}
\end{figure}

In this paper, we consider a set $\mathcal{M}$ of \glspl{vBS} belonging to the same \gls{MNO} and deployed over the same cell. Each \gls{vBS} is tailored to meet the performance requirements of a specific \gls{uRLLC} service. The implementation of these \glspl{vBS} is illustrated in Fig.~\ref{fig:FrameworkIntegration}, featuring both dedicated and shared components. Specifically, the \gls{CU}-\gls{CP} and the \gls{RU} are shared among all \glspl{vBS}~\cite{Ordonez-Lucena2021}, whereas the \gls{CU}-\gls{UP} and the \gls{DU} are dedicated to each \gls{vBS}~\cite{Chang2018}. 

The possibility of sharing \gls{CU}-\gls{CP} in O-RAN is crucial as to enable the centralized execution of \gls{RRM} and \gls{RRC} tasks across \gls{uRLLC} services, preventing conflicting decisions and enhancing overall performances and scalability. Focusing on the \gls{RB} allocation, we consider two hierarchical allocation levels. At the higher level, \glspl{RB} are distributed centrally among different \gls{vBS}-\glspl{DU}, ensuring coordinated resource assignment across services. At the lower level, each \gls{vBS}-\gls{DU} assigns its allocated \glspl{RB} to its attached \glspl{UE} at \gls{MAC} level. Given the necessity of \gls{RT} orchestration decisions for the high-level allocation, this process must occur close to the \gls{vBS}-\glspl{DU}. Specifically, we assume the high-level allocation is executed within the \gls{CU}-\gls{CP}, which ensures that \glspl{RB} are assigned in a coordinated manner to all running services, improving individual service performance and preventing adverse impacts.


Under this scenario, we address the dynamic \gls{RB} allocation problem among multiple \gls{uRLLC} services, each with specific performance requirements regarding packet delay budget and violation probability. We focus on the operation of both the near-\gls{RT} control loop and a new \gls{RT} control loop. To achieve this, we develop a novel framework, namely \name{}, which leverages the \gls{O-RAN} architecture~\cite{O-RAN-workflow,O-RAN-workflow2} and operates at different time scales, as illustrated in Fig.~\ref{fig:FrameworkIntegration}. Next, we explain how \name{} operates at near-\gls{RT} and \gls{RT} scales, followed by details of the \name{}'s components. 

\begin{figure}[t!]
    \centering
    \includegraphics[width=\columnwidth]{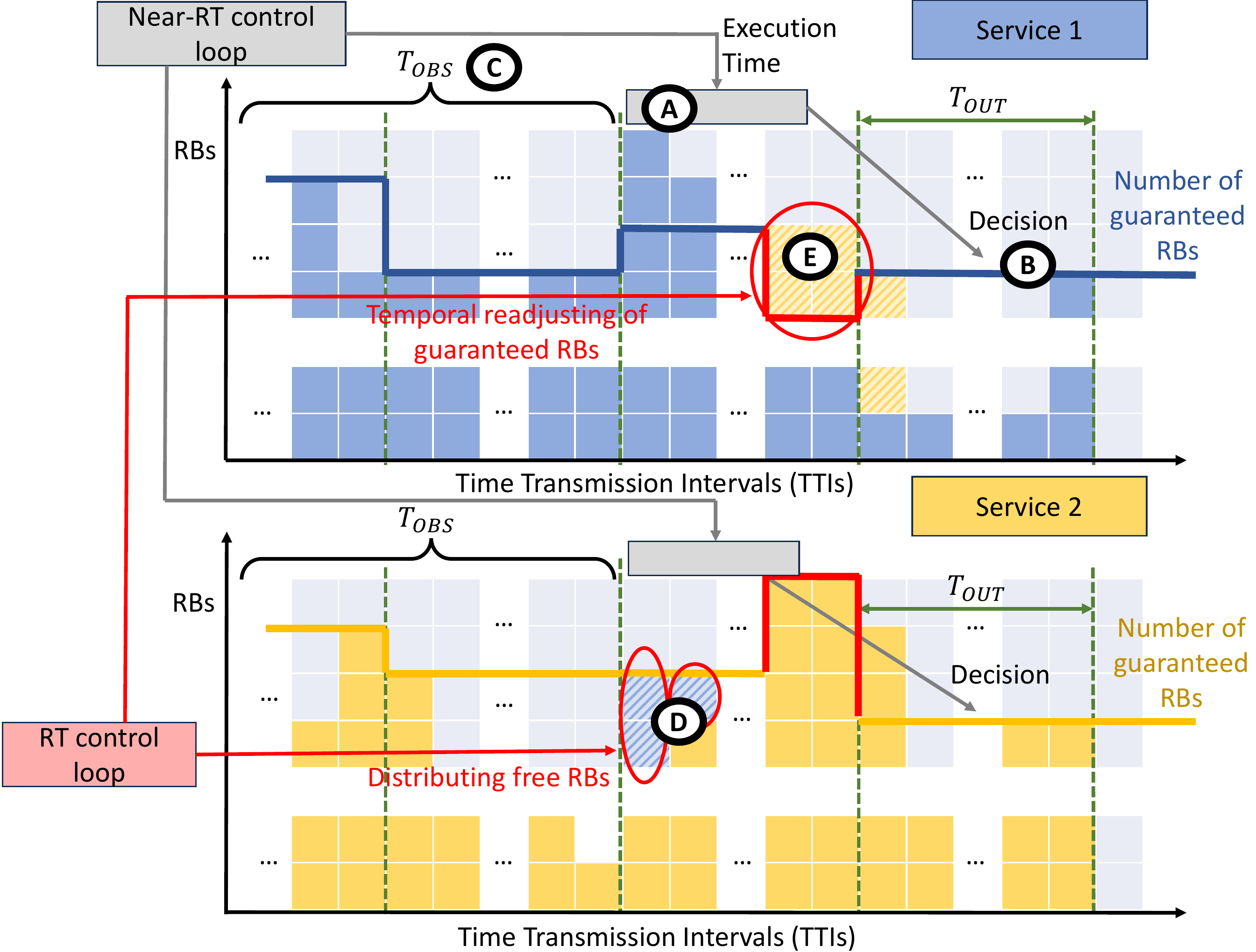}
    \caption{Illustrative example of how \name{} performs dynamic RB allocation at near-\gls{RT} and \gls{RT} scales.}
    \label{fig:ExampleFramework}
\end{figure}

Fig.~\ref{fig:ExampleFramework} showcases near-\gls{RT} and  \gls{RT} control loops. At the near-\gls{RT} scale, \name{} runs periodically every $T_{OUT}$ \glspl{TTI}(see point A). During each execution period, \name{} computes the amount of guaranteed \glspl{RB} for each \gls{uRLLC} service related to the following execution period (see point B). This calculation requires performance metrics from the previous $T_{OBS}$ \glspl{TTI} (see point C). At \gls{RT} scale, within a specific \gls{TTI}, some services may not fully utilize their guaranteed \glspl{RB}, while others may require additional \glspl{RB} beyond what is guaranteed. To optimize the use of available \glspl{RB} in the cell, \name{} implements a \gls{RT} control loop to redistribute previously assigned \glspl{RB} to services that need them (see point D). Furthermore, if traffic anomalies are detected, the \gls{RT} control loop may temporarily adjust the amount of guaranteed \glspl{RB} to meet service requirements (see point E).


To perform near-\gls{RT} and \gls{RT} control loops,  \name{} comprises four \emph{xApps} and one \emph{dApp}\footnote{The concept of \emph{dApps}, introduced in~\cite{dApps_article}, is similar to \emph{rApps} and \emph{xApps}, and enables fine-grained real-time control tasks. These \emph{dApps} can be deployed within the \gls{CU}-\gls{CP},  \gls{CU}-\gls{UP}, and/or \gls{DU}.} as shown in Fig.~\ref{fig:FrameworkIntegration}. The Cell Capacity Estimator, Traffic Estimator, \gls{RB} utilization Estimator and Delay-Aware \gls{RB} Allocation Controller \emph{xApps} are located in the near-\gls{RT} \gls{RIC} and are responsible for computing the guaranteed \glspl{RB} for multiple \gls{uRLLC} services in a near-\gls{RT} scale. The \gls{RT} Controller \emph{dApp} is located in a \gls{CU}-\gls{CP} and it is responsible for fine-tuning \gls{RB} allocation at \gls{RT} scale. Details about these \emph{Apps} are provided below.

\textbf{Traffic Estimator \emph{xApp}}: 
It analyzes the traffic generated by each service $m \in \mathcal{M}$ via the \texttt{E2} interface over the last $T_{OBS}$ \glspl{TTI}, specifically tracking the number of bits generated by each service in each \gls{TTI}. Using this data, this \emph{xApp} estimates the \gls{PMF} of traffic generation for each service. Further details are provided in  in Section~\ref{sec:uRLLCTrafficModel}.  

\color{black}
The \gls{O-RAN} E2 Service Model (E2SM)~\cite{o-ran2} defines a metric to capture the number of bits entering the \gls{RLC} layer in the downlink for a single \gls{DRB}. This metric, measured in Kbits, can be filtered per network slice (and thus, per service) and measured periodically over a period referred to as \textit{granularity period}. This period can be as short as 1 ms. The \texttt{E2} interface does not need to transmit measurements of this metric every \gls{TTI}, but can instead send them in bursts. 
The corresponding \textit{reporting period} should be set such that the measurements arrive within the execution period of \name{} (see point A, Fig.~\ref{fig:ExampleFramework}) and allow it to be executed on time. Additionally, for the proposed Traffic Estimator \emph{xApp}, assuming each measurement requires $N_{bit}$ bits to represent the value of such metric and there is one measurement per service $m \in \mathcal{M}$ over $T_{OBS}$ \glspl{TTI}, the proposed \emph{xApp} must allocate a buffer of size $N_{bit} \times |\mathcal{M}| \times T_{OBS}$ to store the metrics captured via the \texttt{E2} interface.
\color{black}

\textbf{\gls{RB} utilization Estimator \emph{xApp}}:  Assuming a specific amount of guaranteed \glspl{RB} for a service $m\in \mathcal{M}$, it estimates the \gls{PMF}  for the number of \glspl{RB} that the service $m$ uses beyond the guaranteed amount in an arbitrary \gls{TTI}. To achieve this, it utilizes a \gls{NN} based on \gls{MDN}, considering the following inputs: (a) the incoming traffic demand in each \gls{TTI} expressed in bits, (b) the enqueued bits in each \gls{TTI} and (c) the candidate number of guaranteed \glspl{RB} for the next $T_{ OUT}$ \glspl{TTI}. These metrics are available from \texttt{E2} interface. Details about the \gls{MDN} implementation are provided in Section~\ref{sec:NN_rb_util}.

\color{black}
The \gls{O-RAN} Near-Real-time RAN Intelligent Controller E2 Service Model (E2SM) specification \cite{o-ran3} defines the concrete metrics to obtain inputs (b) and (c) through the \texttt{E2} interface. Note that input (a) was previously explained in the context of the Traffic Estimator \emph{xApp}. Input (b), the enqueued bits in each \gls{TTI}, can be obtained from the metric \textit{DL Buffer Occupancy} at the \gls{RLC} layer, which is measured in Kbytes. Input (c), the candidate number of guaranteed \glspl{RB} for the next $T_{ OUT}$ \glspl{TTI}, can be obtained from the parameter \textit{Slice PRB Quota}, which specifies, among other metrics, the number of guaranteed \glspl{RB} allocated to each slice, and thus per service.
\color{black}

\textbf{Cell Capacity Estimator \emph{xApp}}:  It analyzes the packets transmitted via the radio interface as well as the \glspl{MCS} used for these transmissions for each service $m\in \mathcal{M}$ during the last $T_{OBS}$ \glspl{TTI}. This data is collected from the \gls{vBS}-\glspl{DU} via \texttt{E2} interface, \textcolor{black}{as defined in the O-RAN specifications~\cite{o-ran3}. The \glspl{MCS} are computed by the \gls{vBS}-\glspl{DU} using the \gls{UE}'s \gls{CSI} reporting, which includes the \gls{CQI}~\cite{Markova2024}. These measurements are aligned with the metrics specified in 3GPP TS 28.552 (Section 5.1.1.12)~\cite{3gpp_ts_28_552_v19_2_0}.} By considering these measurements, along with a guaranteed amount of \gls{RB} and the \gls{PMF} of potential \glspl{RB} usage exceeding this guaranteed amount (provided by the \gls{RB} utilization Estimator \emph{xApp}), this \emph{xApp} generates the \gls{PMF} for the served traffic. Specifically, it calculates the \gls{PMF} of the number of bits served per \gls{TTI} for each service $m \in \mathcal{M}$. Details on computing these \glspl{PMF} are provided in Section~\ref{sec:CellCapacityEstimator}.

\textbf{Delay-Aware \gls{RB} Allocation Controller \emph{xApp}}: Using inputs from previous \emph{xApps} and an iterative process, this \emph{xApp} calculates the amount of guaranteed \glspl{RB} $N_{ m}^{ min}$ for each service $m$ $\in \mathcal{M}$, which ensures $\mathbb{P}\left[w>W_{ m}\right]<\varepsilon_{ m}$, where $w$ is the packet transmission delay\footnote{\emph{Packet transmission delay} is the waiting time of a Transport Block (TB) unit from entering the transmission buffer until it is fully transmitted.}, $W_{ m}$ is a delay bound and $\varepsilon_{ m}$ is a violation probability. Note that $\sum_{ m \in \mathcal{M}}N_{ m}^{ min}  \leq N_{ cell}^{ RB}$, where $N_{ cell}^{ RB}$ is the \glspl{RB} available in the cell. 

In our previous work~\cite{Adamuz2024} we make use of an \gls{SNC}-based model to derive this probability. While this approach ensures conservative estimations of the delay probabilities, it may limit the number of services accommodated in the cell. To address this drawback, in this paper, we consider a complementary Martingales-based approach that achieves more accurate delay estimations, However, unlike the \gls{SNC}-based model, such estimations are no longer conservative. This translates into the possibility of admitting more services given a more efficient resource allocation $N_{ m}^{ min}$ for each service $m$, at the cost of a relaxed delay probability characterized by a small margin of error. We envision an interested \gls{MNO} selecting the most appropriate model—whether the \gls{SNC}-based or Martingales-based—depending on its business requirements.
Section~\ref{sec:MartingaleTheory} describes the Martingales-based model, while the implementation of this \emph{xApp} is detailed in Section~\ref{sec:ControlLoopRBAllocation}. A comparison of performances of using both the \gls{SNC}-based model and the Martingales-based model is evaluated in Section~\ref{sec:PerformanceResults}.

\textbf{\gls{RT} Controller \emph{dApp}}: It operates every \gls{TTI} and ensures that each \gls{DU} MAC scheduler, one per service $m$ $\in \mathcal{M}$, has available at least $N_{ m}^{ min}$ \glspl{RB}, i.e., the amount computed by the Delay-Aware \gls{RB} Allocation Controller \emph{xApp}. For a given \gls{TTI} $t$, if a service $m$ requires $N_{ t}$ \glspl{RB} such as $N_{ t}<N_{ m}^{ min}$, this \emph{dApp} allocates $N_{ t}$ \glspl{RB} to such service. Otherwise, it first checks for free \glspl{RB}, i.e., those allocated to other services but unused in the current \gls{TTI}. If free \glspl{RB} are found, the \emph{dApp} allocates $N_{ m}^{ min}$ plus the available free \glspl{RB} to the specified service. Additionally, this \emph{dApp} monitors the transmission buffers for traffic anomalies and, if detected, temporarily updates $N_{ m}^{ min}$ $\forall m \in \mathcal{M}$ to mitigate the violation probability. To effectively monitor metrics at \gls{RT} scale, \gls{vBS} components and \emph{dApps} need interfaces similar to those used by \glspl{RIC} and \gls{vBS} components (i.e., \texttt{O1} and \texttt{E2} interfaces). Northbound interfaces between \emph{dApps} and the near-\gls{RT} \gls{RIC}, and southbound interfaces between \emph{dApps} and programmable functions of \glspl{DU}/\gls{CU} as proposed in~\cite{dApps_article}, facilitate control and data sharing, ensuring \emph{dApps} are platform-independent and interact seamlessly with other \gls{O-RAN} components. 

The \gls{O-RAN} Alliance has not yet released specific guidelines detailing how these northbound and southbound interfaces should be implemented. Nevertheless, progress is being made in this direction, as evidenced by a recent \gls{O-RAN} research report~\cite{O-RANdAppReport}. This report outlines the minimum architectural and interface requirements for \emph{dApps}, including data exchange timelines, and explores their impact on control and user plane extensions.

\textbf{\name{} in a Multi-Cell Scenario}: To extend the proposed \name{} framework to a multi-cell scenario, it is sufficient to deploy our solution in a distributed manner. Specifically, one instance of \name{} can be deployed in each network cell. Each instance works independently using local performance metrics from its own cell, allowing the framework to scale efficiently as more cells are added. This setup ensures that our solution remains effective and scalable without needing major changes to its design or operation.

\section{Latency Model Based on Martingales}\label{sec:MartingaleTheory}
To assess the packet transmission delay for each service, we propose a delay model based on Martingales. This section first provides some fundamentals on Martingales. Then, we describe the steps to compute the delay bound $W_{m}$ $\forall m \in \mathcal{M}$. Finally, we particularize these steps to our scenario.

\subsection{Fundamentals on Martingales}
\textit{Definition 1 (Martingale)}~\cite{Yu2024}: Let $(\Omega,{\cal F_{\infty }},P)$ be a probability space and let $\{ \mathcal{F}_{t}, t \ge 0 \}$ be a filtration, which is a nondecreasing sequence of $\sigma$-fields of ${\mathcal{F}_{\infty}},{\mathcal{F}_{t}} \subset {\mathcal{F}_{t + 1}},t \ge 0$. A stochastic process $\{ X(t),t \ge 0 \}$  is adapted to the filtration $\{ \mathcal{F}_{t}, t \ge 0 \}$ if $X(t)$ is $\mathcal{F}_{t}$-measurable for all $t \ge 0$. If $\mathbb{E}[|X(t)|] < \infty\; \forall t\ge 0$ and $\mathbb{E}[X(t + 1)|{ {\mathcal{F}_{t}}}] = X(t)\; \forall t\ge 0$, then $\{ X(t),\;t \ge 0 \}$ is called a martingale. Moreover, if $\mathbb{E}[X(t + 1)|{ {\mathcal{F}_{t}}}] \ge X(t)$, then $\{ X(t),t \ge 0\}$ is called a submartingale. On the contrary, if $\mathbb{E}[X(t + 1) \mid \mathcal{F}_{t}]< X(t)$, then $\{ X(t),t \ge 0\}$ is called a supermartingale.

Let $A(i,j)=\sum_{k=i+1}^{j}a_k$ denote the arrival process, representing the cumulative number of bits arriving at a network node over the time interval $(i,j]$, where $i$ and $j$ represent the \gls{TTI} indexes. Furthermore, $S(i,j)=\sum_{k=i+1}^{j}s_k$ denotes the service process, representing the cumulative number of bits that this network node can process in the same interval. Both processes are driven by the stochastic processes $a_k$ and $s_k$, representing the number of incoming and served bits in an arbitrary \gls{TTI}, respectively. Henceforth, we define $ A(j) := A(0, j) $ and $ S(j) := S(0, j) $ to simplify notation.



\textit{Definition 2 (Arrival Martingale)}~\cite{Poloczek2015}: The arrival process $A(j)$ admits an arrival martingale if for a free parameter $\theta > 0$ there is a $K_{a}\geq 0$ and a function $h_{a}:\text{rng}(a)\rightarrow \mathbb{R}^{+}$ such that the following process is a supermartingale
\begin{equation}
h_{a}\left(a_j\right)\text{exp}\left[\theta(A(j)-jK_{a})\right],\; j\geq 0.
\label{eq:ArrivalMartingale}
\end{equation}
where ‘rng’ represents the range operator. The parameters $K_{a}$ and $h_{a}$ implicitly depend on $\theta$; for brevity, the notation $K_{a}\left(\theta\right)$ and $h_{a}\left(\theta\right)$ is omitted.

\textit{Definition 3: (Service Martingale)}~\cite{Poloczek2015}: The service process $S(j)$ admits a service martingale if for every $\theta > 0$ there is a $K_{s}\geq 0$ and a function $h_{s}:\text{rng}(s)\rightarrow \mathbb{R}^{+}$ such that the next process is a supermartingale
\begin{equation}
h_{s}\left(s_j\right)\text{exp}\left[\theta(jK_{s}-S(j))\right],\; j\geq 0.
\label{eq:ServiceMartingale}
\end{equation}




Assuming the arrival process $A(j)$ and the service process $S(j)$ admit arrival and service martingales, respectively, the probability the packet's transmission delay $w$ exceeds a delay bound $W$, i.e., the \emph{violation probability}, has been defined by Ciucu et al.~\cite{Ciucu2014, Poloczek2014} as 
\begin{equation} 
\mathbb{P}\left[w\geq W\right]\leq\frac{\mathbb{E}\left[h_{a}\left(a_{0}\right)\right]\mathbb{E}\left[h_{s}\left(s_{0}\right)\right]}{H}\text{exp}\left[-\theta^{\ast}K_{s}W\right], 
\label{eq:Violation Probability}
\end{equation}
where $\theta^{\ast}$ is computed as the supremum of the following set
\begin{equation}
\theta^{\ast}:=\sup\left\{\theta > 0:K_{a}\leq K_{s}\right\}, 
\label{eq:thetaoptimum}
\end{equation}
and $H$ is the minimum value of the next set
\begin{equation}
H:=\min\left\{h_{a}\left(x\right)h_{s}\left(y\right): x-y > 0\right\}. 
\end{equation}

In Eq.~\eqref{eq:Violation Probability}, the delay bound $W$ is given in terms of the number of \glspl{TTI}. Based on this, we can reformulate Eq.~\eqref{eq:Violation Probability} to estimate the delay bound $W$ as shown in Eq.~\eqref{eq:DelayBoundTheory}. Note that we have multiplied the resulting expression by the duration of a single \gls{TTI}, $t_{slot}$, to convert the resulting delay bound into time units.
\begin{equation}
W \leq -\text{log}\left[\frac{H\mathbb{P}\left[w\geq W\right]}{\mathbb{E}\left[h_{a}\left(a_{0}\right)\right]\mathbb{E}\left[h_{s}\left(s_{0}\right)\right]}\right] \left(\theta^{\ast}K_{s}\right)^{-1} t_{slot}.
\label{eq:DelayBoundTheory}
\end{equation}

\subsection{Methodology for Delay Bound Computation with Martingales}
Based on the fundamentals of Martingales discussed earlier, the following steps outline the procedure for estimating the delay bound $W$ given a target violation probability $\mathbb{P}\left[w\geq W\right]$.

\textbf{Step 1}: Define the arrival and service martingales for the arrival and service processes $A(j)$ and $S(j)$, respectively. In this paper, we use Wald’s martingales, which are applied to sums of i.i.d. random variables~\cite{Wald1944}. 

\textit{Definition 4 (Wald's Martingale)~\cite{Wald1944}:} Let $\{x_j,j >1\}$ be a sequence of i.i.d. random variables with a finite \gls{MGF} $M_x\left(\theta\right)= \mathbb{E}\left[e^{\theta x}\right] $ for some $\theta > 0$, and let $Y(j) = \sum_{k=1}^{j} x_k$ with $Y(0) = 0$. The process $M_Y^W(j)$ $\forall j \geq 0$ , defined by $M_Y^W(j) = M_x\left(\theta\right)^{-j} e^{\theta Y(j)}$, is Wald’s martingale, satisfying $\mathbb{E}\left[M_Y^W(j)\right] = 1$ $\forall j \geq 0$.

\textbf{Step 2}: Rewrite Wald’s martingales to conform with the arrival and service martingales as specified in Eqs.~\eqref{eq:ArrivalMartingale} and~\eqref{eq:ServiceMartingale}. From these reformulated expressions, extract the parameters  $h_{a}\left(a_j\right)$, $K_{a}$, $h_{s}\left(s_j\right)$ and $K_{s}$.

\textbf{Step 3}: Using the extracted parameters, compute the threshold $H$, the expectations $\mathbb{E}\left[h_{a}\left(a_{0}\right)\right]$, $\mathbb{E}\left[h_{s}\left(s_{0}\right)\right]$, and $\theta^{\ast}$.

\textbf{Step 4}:  Finally, substitute the computed values into Eq.~\eqref{eq:DelayBoundTheory} to estimate the delay bound $W$. 

Below, we particularize these steps to estimate the delay bound $W_m$ for a \gls{uRLLC} service $m \in \mathcal{M}$.

\subsection{Arrival Process of a Downlink uRLLC Service}\label{sec:uRLLCTrafficModel}
The arrival process $a_{ m,j}$ represents the number of bits that arrive to the \gls{vBS}'s transmission buffer for service $m$ in the $j$-th \gls{TTI}. We assume the \gls{PMF} of the process $a_{ m,j}$ can be estimated by using samples of the incoming bits per \gls{TTI} in the last $T_{ OBS}$ \glspl{TTI}. To that end, we define the sample vector $\vec{x}_{ a_{ m}} =\{a_{ m,1}^{ in}, a_{ m,2}^{ in}\, \hdots\, a_{ m,T_{ OBS}}^{ in} \}$, where $a_{ m,i}^{ in}$ denotes the number of bits that arrived to the transmission buffer in the \gls{TTI} $i$ for the service $m$. Additionally, $a_{ m,i}^{ in}=\sum_{k=1}^{J_{ m,i}^{ in}}l_k$, where $J_{ m,i}^{ in}$ is the number of incoming packets for service $m$ in the \gls{TTI} $i$ and $l_{ k}$ the size of the packet $k$. Note that the computation of $\vec{x}_{ a_{ m}}$ is a task performed by the Traffic Estimator \emph{xApp}.
Under these assumptions, we can compute the \gls{MGF} for the process $a_{ m,j}$ as
\begin{equation}
    M_{ a_{ m}}\left(\theta\right) = \left(T_{ OBS}\right)^{-1}\sum_{i=1}^{T_{ OBS}} \text{exp}\left[\theta a_{ m,i}^{ in}\right].
    \label{eq:MGFTraffic}
\end{equation}
Based on the previous equation, we can build the Wald's martingale for the arrival process as 
\begin{equation}
\begin{split}
    M_{ a_{ m}}^{W}\left(\theta\right) & = M_{ a_{ m}}\left(\theta\right)^{-j}\text{exp}\left[\theta A(j)\right] \\
    & = \text{exp}\left[-j\text{log}\left[ M_{ a_{ m}}(\theta)\right]\right]\text{exp}\left[\theta A(j)\right] \\
    & = \text{exp}\left[\theta\left(A(j) - j\theta^{-1}\text{log}\left[ M_{ a_{ m}}(\theta)\right]\right)\right].\\
\end{split}
    \label{eq:WaldTraffic}
\end{equation}
%
Note that the final expression was obtained by applying the exponential operator to the logarithm of the term $M_{ a_{ m}}\left(\theta\right)^{-j}$, and then operating on the resulting expression.
The obtained Wald's martingale has the same form of the arrival-martingale defined in Eq.~\eqref{eq:ArrivalMartingale}. Based on this observation, 
we can obtain the parameter $K_{a,m}$ as 
\begin{equation}
K_{a,m} = \theta^{-1}\text{log}\left[\left(T_{ OBS}\right)^{-1}\sum_{i=1}^{T_{ OBS}} \text{exp}\left[\theta a_{ m,i}^{ in}\right]\right].
    \label{eq:Ka}
\end{equation}
Note that the parameter $h_{a,m}(a_{m,j})=1$. Without loss of generality, 
we define the auxiliary parameter $K_{a,m}^{\prime} = \theta K_{a,m}$.

\subsection{Service Process of the Cell Capacity Provided to a Downlink uRLLC Service}\label{sec:ServiceModel}
The service process $s_{ m,j}$ represents the number of bits that may be served by the cell for service $m \in \mathcal{M}$ in the $j$-th \gls{TTI}. The negative \gls{MGF} for $s_{m,j}$ is defined in Eq. \eqref{eq:MGFAvailableCapacityforSlice}. To that end, we consider the \gls{PMF} of $s_{m,j}$ can be estimated by (a) using samples of the number of bits which may be transmitted in the last $T_{ OBS}$ \glspl{TTI} and (b) considering the \gls{PMF} of the \gls{RB} utilization. The latter captures the dynamics of the \gls{RT} Controller \emph{dApp}. Specifically, we consider the sample vectors $\vec{x}_{s_{m},n}  =\{s_{m,1}^{ n,out}, s_{m,2}^{ n,out}\, \hdots\, s_{m,T_{ m,n}}^{ n,out} \}$  $\forall n \in [0,\; 
N_{ add}]$ where $s_{m,i}^{ n,out}$ denotes the number of bits which may be transmitted by the cell for service $m$ in a single \gls{TTI}, i.e., considering the serving cell uses $n + N_{ m}^{ min}$ \glspl{RB} for such service. Note that $n$ represents the number of additional \glspl{RB} allocated to service $m$, i.e, beyond the guaranteed ones. Furthermore, $N_{ add} = N_{ cell}^{ RB}-N_{ m}^{ min}$ and $T_{ m,n}$ is the number of samples. Additionally, we consider $\pi_{ m,n}$ as the probability that the service $m$ has a number of available \glspl{RB} $n+N_{ m}^{ min}$ in an arbitrary \gls{TTI}, conditioned to the fact that the service $m$ requires $n$ \glspl{RB} more than $N_{ m}^{ min}$. The computation of $\vec{x}_{ s_{m},n}$, a task performed by the Cell Capacity Estimator \emph{xApp}, is detailed in Section \ref{sec:CellCapacityEstimator}. 
\begin{equation}
    M_{ s_{m}}(- \theta) =\sum_{n=0}^{
N_{ add}} \frac{\pi_{ m,n}}{T_{ m,n}} \sum_{i=1}^{T_{ m,n}} \text{exp} \left[ - \theta s_{m,i}^{ n,out}\right].
\label{eq:MGFAvailableCapacityforSlice}
\end{equation}

Based on the previous equation, we can derive the Wald's martingale for the service process as shown in Eq.~\eqref{eq:WaldService}. Note that the final expression was obtained by applying the exponential operator to the logarithm of the term $M_{ s_{ m}}\left(-\theta\right)^{-j}$.
\begin{equation}
\begin{split}
    M_{ s_{ m}}^{W}(-\theta) & = M_{ s_{ m}}\left(-\theta\right)^{-j}\text{exp}\left[-\theta S(j)\right] \\
     & = \text{exp}\left[-j\text{log}\left[ M_{ s_{ m}}(-\theta)\right]\right]\text{exp}\left[-\theta S(j)\right] \\
    & = \text{exp}\left[\theta\left(- j \theta^{-1}\text{log}\left[ M_{ s_{ m}}(-\theta)\right]-S(j)\right) \right].\\
\end{split}
    \label{eq:WaldService}
\end{equation}

The resulting Wald's martingale has the same form as the service martingale defined in Eq.~\eqref{eq:ServiceMartingale}. Based on that, we can obtain the parameter $K_{s,m}$ as Eq.~\eqref{eq:Ks} shown. Note that the parameter $h_{s,m}(s_{m,j})=1$. For convenience, we define the auxiliary parameter $K_{s,m}^{\prime} = \theta K_{s,m}$.
\begin{equation}
K_{s,m} = -\left(\theta^{-1}\right)\text{log}\left[\sum_{n=0}^{
N_{ add}} \frac{\pi_{ m,n}}{T_{ m,n}} \sum_{i=1}^{T_{ m,n}} \text{exp} \left[ - \theta s_{m,i}^{ n,out}\right]\right]
    \label{eq:Ks}
\end{equation}

\subsection{Delay Bound Estimation for an uRLLC Service}\label{sec:DelayBoundEstimation}
Since $h_{a,m}(a_{m,j})=1$ and $h_{s,m}(s_{m,j})=1$, the threshold $H_m=1$ as well as the expectations $\mathbb{E}\left[h_{a,m}\left(a_{m,0}\right)\right]=1$ and $\mathbb{E}\left[h_{s,m}\left(s_{m,0}\right)\right]=1$. Using these results along with Eq.~\eqref{eq:DelayBoundTheory} and Eq.~\eqref{eq:Ks}, we can define $W_{ m}$ $\forall m \in \mathcal{M}$ as follows
\begin{equation}
\begin{split}
    W_m & \approx \frac{\text{log}\left[\mathbb{P}\left[w\geq W_m\right]\right]}{\text{log}\left[\sum_{n=0}^{
N_{ add}} \frac{\pi_{ m,n}}{T_{ m,n}} \sum_{i=1}^{T_{ m,n}} \text{exp} \left[ - \theta_{m}^{\ast} s_{m,i}^{ n,out}\right]\right]}\\
\end{split}
\label{eq:DelayBoundEstimation}
\end{equation}

The previous delay bound depends on $\theta_{m}^{\ast}$, which is the minimum upper bound for the set of values of $\theta$ where $K_{s,m} \geq K_{a,m}$ as Eq. \eqref{eq:thetaoptimum} shown. Note this set of values is the same when considering the inequality $K_{s,m}^{\prime} \geq K_{a,m}^{\prime}$. 

\begin{figure}[t!]
    \centering
    \includegraphics[width=\columnwidth]{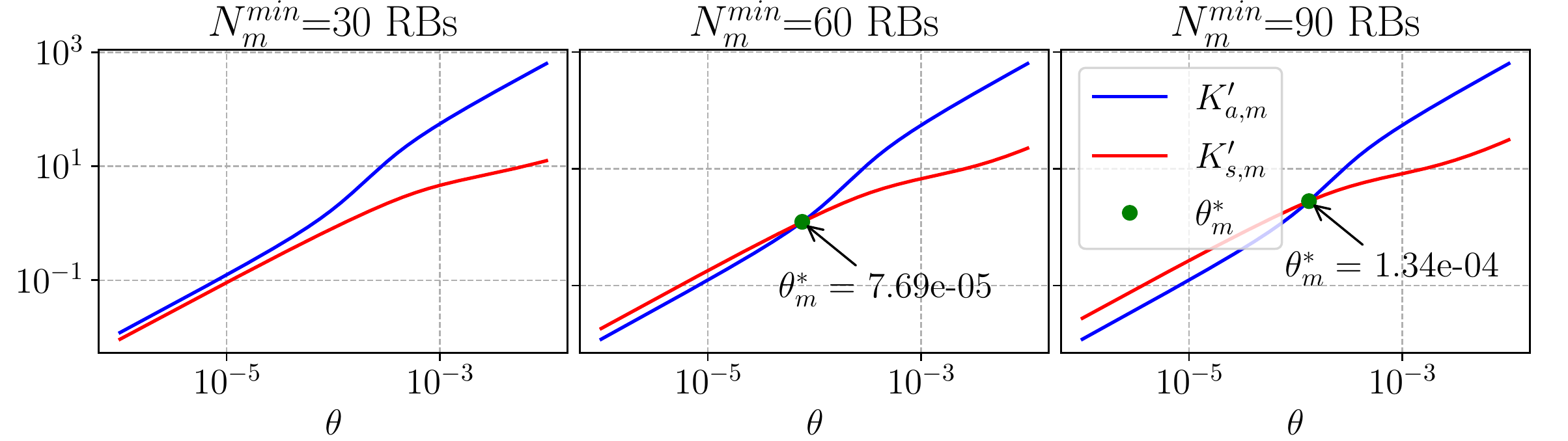}
    \caption{Evaluation of $K_{a,m}^{\prime}$ and $K_{s,m}^{\prime}$ for a service $m \in \mathcal{M}$ with a specific number of guaranteed \glspl{RB} $N_m^{min}$. This example uses the traffic pattern and channel conditions defined in the experimental setup (see Section~\ref{sec:PerformanceResults}). }
    \label{fig:ValidationKaKs}
\end{figure}

Before computing $\theta_{m}^{\ast}$, we first present an example in Fig.~\ref{fig:ValidationKaKs}, illustrating the evolution of the parameters $K_{a,m}^{\prime}$ and $K_{s,m}^{\prime}$ for various $\theta$ values. This example consider different scenarios, each with a specific number of guaranteed \glspl{RB} for a single service $m \in \mathcal{M}$.  If we examine the right plot, i.e., $N_{ m}^{ min}=90$ \glspl{RB}, $K_{s,m}^{\prime}$ is greater than $K_{a,m}^{\prime}$ for very small $\theta$ values. The range where $K_{s,m}^{\prime} > K_{a,m}^{\prime}$ decreases as the number of guaranteed \glspl{RB} decreases, e.g., when $N_{ m}^{ min}=60$ \glspl{RB}. Furthermore, when the reserved \glspl{RB} are sufficiently small, e.g., $N_{ m}^{ min}=30$, $K_{a,m}^{\prime}$ never exceed $K_{s,m}^{\prime}$ for any $\theta$, indicating that the \gls{DL} traffic for the service cannot be adequately served over time, resulting in overflow in the \gls{vBS}'s transmission buffer.

\begin{algorithm}[!t]
\SetAlgoLined
\small
\textbf{Inputs:} Sample vectors $\vec{x}_{ a_{ m}}$ and $\vec{x}_{s_{m},n}$, $T_{ OBS}$, $T_{ m,n}$, $\pi_{ m,n}$\;
\textbf{Initialization:} $\theta_{ old} = 1$, $\Delta = 0.9$\;
\While{True}{
    Compute $\theta_{ new} = \theta_{old}\Delta$\;
    \eIf{$K_{s,m}^{\prime}(\theta_{ new}) - K_{a,m}^{\prime}(\theta_{ new}) \geq 0$}{
        \textbf{break}\;
    }{
        $\theta_{ old} = \theta_{ new}$\;
        \If{$\theta_{ new} < 10^{-9}$}{
            \textbf{break}\;
        }
        
    }
}
\eIf{$\theta_{ new} \geq 10^{-9}$}{
    Use bisection algorithm~\cite{burden19852} to find $\theta_{m}^{\ast} \in [\theta_{ new}, \theta_{ old}]$\;
    \textbf{Output:} $\theta_{m}^{\ast}$\;
}{
    \textbf{Output:} $\theta_{m}^{\ast}$ does not exist\;
}
\caption{Search of $\theta_{m}^{\ast}$ for a service $m \in \mathcal{M}$.}
\label{alg:Search-for-optimum-theta}
\end{algorithm}

Another observation is $K_{a,m}^{\prime}$ and $K_{s,m}^{\prime}$ are not convex, indicating the need for a heuristic algorithm to search for $\theta_{m}^{\ast}$. Based on the observed behavior of $K_{a,m}^{\prime}$ and $K_{s,m}^{\prime}$, we have proposed the heuristics defined in Algorithm~\ref{alg:Search-for-optimum-theta}. It begins by taking as input the sample vectors $\vec{x}_{ a_{ m}}$, $\vec{x}_{s_{m},n}$ containing the traffic estimation for service $m$ and the cell capacity for that service, respectively, as well as the sample sizes $T_{ OBS}$, $T_{ m,n}$ and the probability $\pi_{m,n}$ (line 1). With these inputs, an initial value is set for the upper limit of the interval where $\theta_{m}^{\ast} \in [\theta_{ new}, \theta_{ old}]$, and a parameter $\Delta$ is defined to determine the speed of the search for $\theta_{m}^{\ast}$, i.e., the closer $\Delta$ is to 1, the more exhaustive the search (line 2). Based on experimental observations in Fig. \ref{fig:ValidationKaKs}, we set $\theta_{old} = 1$ and $\Delta = 0.9$. Then, we start an iterative procedure (lines 3-13). In each iteration, a value for the lower limit of the interval $[\theta_{ new}, \theta_{ old}]$ is computed (line 4). Then, we evaluate if $K_{s,m}^{\prime}(\theta_{new}) - K_{a,m}^{\prime}(\theta_{new}) >0$. If so, the loop terminates with $\theta_{new}$ as the new lower bound (line 6). Otherwise $\theta_{old}$ is updated to $\theta_{new}$ and the loop continues (line 8). Note that if $\theta_{new}$ remains above $10^{-9}$ (line 10), the loop ends and the algorithm stop (lines 18). The value $10^{-9}$ has been experimentally established, based on Fig.~\ref{fig:ValidationKaKs}, as sufficiently small to determine that $\theta_{m}^{\ast}$ does not exist due to the cell capacity being inadequate to meet the service demands.  If $\theta_{new}$ is above $10^{-9}$, the bisection method~\cite{burden19852}, which is based on the intermediate value theorem,  is used for searching the value of $\theta_{m}^{\ast} \in  [\theta_{new}, \theta_{old}]$ (lines 15-16).

\color{black}
The computational complexity of Algorithm~\ref{alg:Search-for-optimum-theta} is $\mathcal{O}\left( \left[T_{m,n} \cdot N_{add} + T_{OBS}\right] \cdot N_{\text{max}} \right)$, where $T_{m,n}$, $N_{add}$ and $T_{OBS}$  are input parameters that determine the number of operations performed by the algorithm. In each iteration, the algorithm computes two variables: $K_{s,m}^{\prime}$ and $K_{a,m}^{\prime}$. To calculate $K_{s,m}^{\prime}$, i.e, Eq.~\eqref{eq:Ks}, a double summation is performed over the indices $T_{m,n}$ and $N_{add}$, with each summation involving the calculation of an exponential function, which is assumed to take constant time. This results in a complexity of $\mathcal{O}(T_{m,n} \cdot N_{add})$. For $K_{a,m}^{\prime}$, i.e., Eq.~\eqref{eq:Ka}, a summation over $T_{OBS}$ is performed, also involving the calculation of an exponential function in each term, yielding a complexity of $\mathcal{O}(T_{OBS})$. Since these calculations are independent, their complexities are added, giving $\mathcal{O}(T_{m,n} \cdot N_{add} + T_{OBS})$ per iteration. The algorithm includes a while loop, whose maximum number of iterations is bounded by $ N_{\text{max}}= \left\lceil \frac{\text{log}(10^{-9})}{\text{log}(\Delta)} \right\rceil $, where $\Delta$ is a parameter that controls convergence. Therefore, the total complexity of Algorithm~\ref{alg:Search-for-optimum-theta} is $\mathcal{O}\left( \left[T_{m,n} \cdot N_{add} + T_{OBS}\right] \cdot N_{\text{max}} \right)$, exhibiting a scalable (and computationally affordable) behavior.

\color{black}

\section{Control Loops of \name{}}\label{sec:SNC-basedOrchestrator}
In this section, we detail how \name{} executes the near-\gls{RT} and \gls{RT} control loops. We begin by explaining the computation of cell capacity samples and the estimation of probabilities $\pi_{m,n}$. Next, we describe how the Delay-Aware \gls{RB} Allocation Controller \emph{xApp} determines $N_{m}^{min}$ for all $m \in \mathcal{M}$ in the near-\gls{RT} control loop. Finally, we outline how the \gls{RT} Controller \emph{dApp} mitigates the violation probability.

\subsection{Computation of the Cell Capacity Samples}\label{sec:TrafficEstimator}\label{sec:CellCapacityEstimator}
The Cell Capacity Estimator \emph{xApp} performs the following steps to obtain a sample $s_{ m,i}^{ n,out} \in \vec{x}_{ s_{ m},n}$.

\textbf{Step 1:} For each transmitted packet $j$, it considers (a) the packet size $l_{ j}$, and (b) the amount of \glspl{RB} required to transmit it, i.e., $N_{ pkt,j}$. Note that we are assuming that a unique \gls{MCS} value is adopted to transmit each packet. It means this \emph{xApp} can compute the number of bits transmitted per \gls{RB} as $s_{ RB,j}=l_{ j}/N_{ pkt,j}$. Based on that, it defines a vector $\vec{x}_{ pkt,j} = \{s_{ RB,j},s_{ RB,j}\, \hdots,s_{ RB,j}\}$, where the element $s_{ RB,j}$ is repeated $N_{ pkt,j}$ times.

\textbf{Step 2:} Performing the previous step for all the transmitted packets, this \emph{xApp} obtains a set of vectors $\vec{x}_{ pkt,j}$ $\forall j \in \mathcal{J}_{ OBS}$, where $\mathcal{J}_{ OBS}$ denotes the set of packets transmitted in the last $T_{ OBS}$ \glspl{TTI}. Based on that, this \emph{xApp} defines $\vec{x}_{ con}=\{\vec{x}_{ pkt,1},\vec{x}_{ pkt,2}\,\hdots \vec{x}_{ pkt,|\mathcal{J}_{OBS}|}\}$ as a vector which concatenates each measured vector $\vec{x}_{ pkt,j}$.

\textbf{Step 3:} Based on $\vec{x}_{ con}$, this \emph{xApp} groups its samples in set of $N_{ m,n}^{ set}=n+N_{ m}^{ min}$ consecutive samples. Note that $n\in \left[0, N_{add}\right]$ is the number of additional \glspl{RB} that may be allocated to service $m$ beyond the guaranteed number of \glspl{RB} (see definition in Section~\ref{sec:ServiceModel}). In turn, each group of $N_{ m,n}^{ set}$ consecutive samples defines a vector $\vec{x}_{ m,i}^{ n,out}$, where $i \in [1,T_{ m,n}]$ represents the i-th vector. We define $T_{ m,n}$ as the total number of built vectors. Note we have one vector $\vec{x}_{ m,i}^{ n,out}$ per sample $s_{ m,i}^{ n,out}$, i.e., see Eq.~\eqref{eq:MGFAvailableCapacityforSlice}.

\textbf{Step 4:} Finally, this \emph{xApp} obtains the sample $s_{ m,i}^{ n,out}$ as the sum of all the elements of the vector $\vec{x}_{ m,i}^{ n,out}$, i.e., $s_{ m,i}^{ n,out}=\sum_{z=1}^{N_{ m,n}^{ set}} \vec{x}_{ m,i}^{ n,out}\{z\}$.

Note that this \emph{xApp} generates a sample vector $\vec{x}_{s{m},n}$ for each value of $n$. Thus, steps 3-4 need to be repeated for every $n$. For instance, if a service $m$ has $N_{m}^{min}=10$ guaranteed \glspl{RB} and the cell has $N_{cell}^{RB}=25$, then $n$ varies from $0$ to $15$.

\subsection{{Mixture Density Networks for Estimating the RB Utilization}}
\label{sec:NN_rb_util}
\begin{figure}[t]
\centering
\includegraphics[width=\columnwidth]{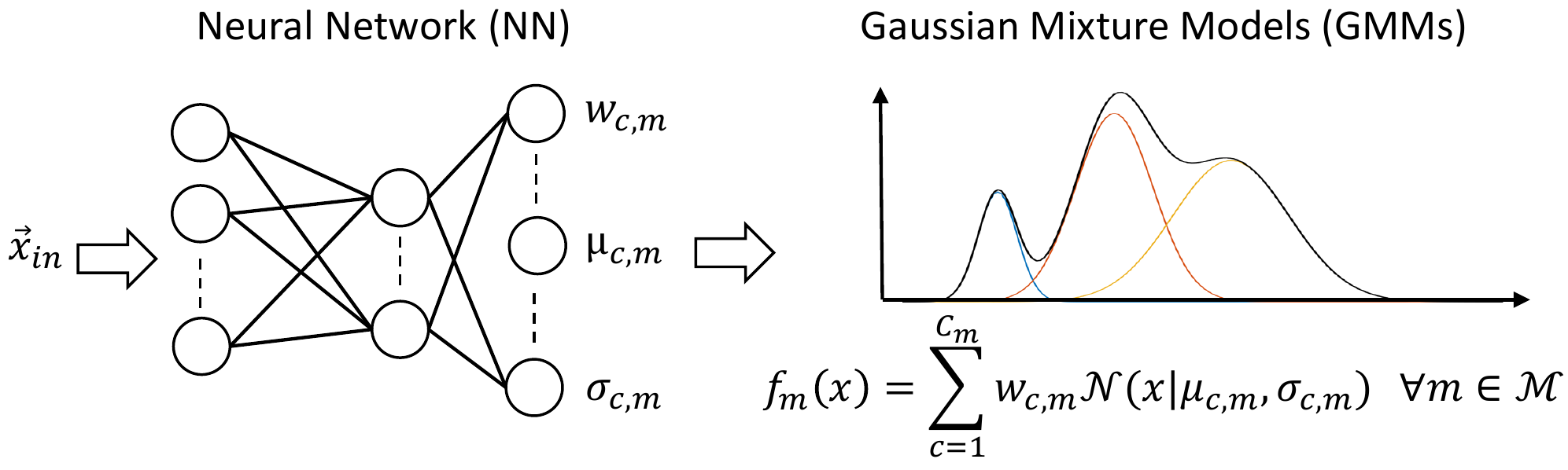}
\caption{Overview of the considered Mixture Density Network (MDN). Copied from \cite{Adamuz2024}.}
\label{fig:MDN}
\end{figure}

The Delay-Aware RB Allocation Controller \emph{xApp} utilizes a \gls{MDN}, a \gls{NN} architecture for modeling probability distributions~\cite{MDN_book}, to estimate $\pi_{m,n}$. Unlike a typical \gls{NN}, which produces a single value, an \gls{MDN} generates one or more parameterized mixture models in its output layer. These models are weighted combination of several component distributions. We consider $|\mathcal{M}|$ \glspl{GMM}, one per service. It is proven the \gls{GMM} accurately approximate any arbitrary distribution in the context of wireless networks~\cite{MDN_Modeling,Gaussian_Mixtures,GMM_basic}. Specifically, $\pi_{m,n}$ might not follow a single known statistical distribution and may vary over time. The \gls{GMM} is described by the equation in Fig.~\ref{fig:MDN}, where $C_{ m}$ denotes the number of Gaussian distributions. In turn, the $c$-th distribution is characterized by the weight $w_{ c,m}$, the mean $\mu_{ c,m}$ and the standard deviation $\sigma_{ c,m}$. Note that $\sum_{c=1}^{C_{ m}}w_{ c,m}=1$.
The considered \gls{MDN} is summarized in Fig.~\ref{fig:MDN}. The inputs parameters $\vec{x}_{ in}$ are: (a) the \gls{RB} utilization for each service, and (b) the 25th, 50th and 75th percentiles of incoming and enqueued bits. All of them are measured in the last $T_{ OUT}$ \glspl{TTI} for each service $m\in \mathcal{M}$. Additionally, the \gls{MDN} considers as input the target number of guaranteed \glspl{RB} for each service $m\in \mathcal{M}$ in the next $T_{OBS}$ \glspl{TTI}. Based on them, the \gls{MDN} provides an estimation of the parameters $w_{ c,m}$, $\mu_{ c,m}$ and $\sigma_{c,m}$. Finally, we compute the $\pi_{ m,n}$ as Eq. \ref{eq:ProbabilitiesSNCCapacity} shows. Specifically, we split the \gls{GMM} into $N_{ add}$ regions (i.e., see Section \ref{sec:ServiceModel}) and compute $\pi_{ m,n} $  as the probability of being in the $n$-th region.
\begin{equation}
    \pi_{ m,n} = \sum_{c=1}^{C_{ m}}\int_{n-0.5}^{n+0.5} w_{ c,m}\mathcal{N}\left(x|\mu_{ c,m},\sigma_{ c,m}\right) dx.
    \label{eq:ProbabilitiesSNCCapacity}
\end{equation}

Note that the considered \gls{MDN} model was validated in our previous work~\cite{Adamuz2024}.

\subsection{\name{} Near-RT Control Loop}\label{sec:ControlLoopRBAllocation}
The Delay-Aware RB Allocation Controller \emph{xApp} aims to determine the optimal allocation of guaranteed \glspl{RB} $N_{ m}^{ min}$ for each service $m \in \mathcal{M}$, such that the delay bound $W_{ m}$ is as close as possible to, or below, the target delay budget $W_{ m}^{th}$, given the target violation probability $\varepsilon_{ m}$. To achieve this, we formulate the \texttt{ORCHESTRATION\_URLLC\_SERVICES} problem, aiming to minimize the maximum ratio $W_{ m}/W_{ m}^{ th}$ across all $|\mathcal{M}|$ services as shown in Eq. \eqref{eq:OptProblem}. This optimization is subject to the constraint  Eq. \eqref{eq:OptORCons1}, ensuring the guaranteed \glspl{RB} across all services does not exceed the available amount given the cell configuration.\\
\noindent \textbf{Problem}~\texttt{ORCHESTRATION\_URLLC\_SERVICES}:
\begin{alignat}{2}
&\underset{N_{ m}^{ min}}{\text{min.}}  \quad &&  g(\vec{W}) = \text{max}\left\{\frac{W_{ 1}}{W_{ 1}^{ th}} ,\hdots, \frac{W_{ |\mathcal{M}|}}{W_{ |\mathcal{M}|}^{ th}}\right\}, \label{eq:OptProblem}\\
&\text{s.t.: } \quad && \sum_{m=1}^{|\mathcal{M}|} N_{ m}^{ min} \leq 
N_{ cell}^{ RB}. \label{eq:OptORCons1}
\end{alignat}

The objective function $g(\vec{W})$ depends on the computation of $W_{ m}$ $\forall m \in \mathcal{M}$. Specifically, for a given number of guaranteed \glspl{RB}  $N_{ m}^{ min}$ $\forall m \in \mathcal{M}$, it is necessary to calculate $|\mathcal{M}|$ delay bounds $W_{m}$. This requires calculating $\theta_{m}^{\ast}$ for each service beforehand. As demonstrated in Fig.~\ref{fig:ValidationKaKs} (Section~\ref{sec:DelayBoundEstimation}), the search for $\theta_{m}^{\ast}$ occurs in a non-convex space, which necessitates using the heuristics from Algorithm~\ref{alg:Search-for-optimum-theta} $|\mathcal{M}|$ times, one per service. For this reason, we propose Algorithm \ref{alg:OptimumOrchestration} to solve the \texttt{ORCHESTRATION\_URLLC\_SERVICES} problem.

\begin{algorithm}[b!]
\SetAlgoLined
\small
\textbf{Inputs:} $W_{ m}^{ th}$, $\varepsilon_{ m}$, $\vec{x}_{ a_{ m}}$, $\vec{x}_{ s_{ m},n}$\;
\textbf{Initialization:} $N_{ m,z}^{ min}=\lfloor 
N_{ cell}^{ RB}/|\mathcal{M}|\rfloor$, $W_{ m}=\infty$, $W_{ m,z}=\infty$ $stop=False$\;
 \While{stop == False}{
 Estimate $\pi_{ m,n}$ $\forall m \in \mathcal{M}$ $\forall n \in [0,
N_{ add}]$ [Section \ref{sec:NN_rb_util}]\;
 Estimate $W_{ m,z}$ $\forall m \in \mathcal{M}$ [See Eq.~\eqref{eq:DelayBoundEstimation}]\;
 Evaluate $g(\vec{W}_{ z})$ and $g(\vec{W})$ [See Eq.~\eqref{eq:OptProblem}]\;
  \eIf{$g(\vec{W}_{ z}) <  g(\vec{W})$ }{
  Update $N_{ m}^{ min} = N_{ m,z}^{ min} $;  $W_{ m} = W_{ m,z}$  $\forall m \in \mathcal{M}$\;
  Select $m^{ \prime} | \frac{W_{ m^{ \prime}}}{W_{ m^{ \prime}}^{ th}} \geq  \frac{W_{ m}}{W_{ m}^{ th}}$ $\forall m \in \mathcal{M} \setminus \{m^{ \prime}\}$\;
  Select $m^{ \prime\prime} | \frac{W_{ m^{ \prime\prime}}}{W_{ m^{ \prime\prime}}^{ th}} \leq  \frac{W_{ m}}{W_{ m}^{ th}}$ $\forall m \in \mathcal{M} \setminus \{m^{ \prime\prime}\}$\;
  Compute $N_{ m^{ \prime},z}^{min} = N_{ m^{ \prime}}^{min} + 1$; $N_{ m^{ \prime\prime},z}^{ min} = N_{ m^{ \prime\prime}}^{min} - 1$\;
   }{
   $stop = True$\;
   }
 }
 return $N_{m}^{min}$, $W_m$\;
 \caption{Near-RT RB allocation}
 \label{alg:OptimumOrchestration}
\end{algorithm}

This algorithm considers the target delay budget $W_{ m}^{ th}$, the target violation probability $\varepsilon_{ m}$ and the sample vectors $\vec{x}_{ a_{ m}}$, $\vec{x}_{ s_{ m},n}$ as inputs. It begins with an equal distribution of the available \glspl{RB} among the services, i.e.,  $N_{ m,z}^{ min}=\lfloor 
N_{ cell}^{ RB}/|\mathcal{M}|\rfloor$. Then, it initiates an iterative procedure to get $N_{ m}^{ min}$ $\forall m \in \mathcal{M}$. In each iteration, considering $N_{ m,z}^{ min}$ guaranteed \glspl{RB} for each service, it first estimates $\pi_{ m,n}$ using the \gls{MDN} described in Section \ref{sec:NN_rb_util} (step 4). Then, it uses the Martingales-based model (see Section \ref{sec:DelayBoundEstimation}) to estimate the delay bound $W_{ m,z}$ for the target $N_{ m,z}^{ min}$ (step 5). Then, it evaluates the objective function, i.e., Eq. \eqref{eq:OptProblem}, considering $W_{ m,z}$ and $W_{ m}$ (step 6). If the objective function has been reduced 
compared to the previous iteration (step 7), the algorithm updates the new values for $N_{ m}^{ min}$ and $W_{ m}$ (step 8) and tries to reduce the objective function. To that end, it selects the service $m^{ \prime}$ with the best ratio $W_{m^{ \prime}}/W_{ m^{ \prime}}^{ th}$ and the service $m^{ \prime\prime}$ with the worst ratio $W_{m^{ \prime\prime}}/W_{ m^{ \prime\prime}}^{ th}$ (steps 9-10). The algorithm then reallocates one guaranteed \gls{RB} from service $m^{\prime\prime}$ to service $m^{\prime}$ (step 11). A new iteration of the algorithm starts if $N_{ m,z}^{ min}$ $\forall m \in \mathcal{M}$ improves the objective function. Conversely, the iterative procedure ends if the objective function can not be further improved (step 13).

\color{black}
Regarding the computational complexity of Algorithm~\ref{alg:OptimumOrchestration}, the bottleneck lies in line 5, which depends on the execution of Algorithm~\ref{alg:Search-for-optimum-theta}. The time complexity results in $\mathcal{O}\left([T_{m,n} \cdot N_{add} + T_{OBS}] \cdot N_{max} \cdot |\mathcal{M}|\right)$, which corresponds to the complexity of Algorithm~\ref{alg:Search-for-optimum-theta} multiplied by the number of services $|\mathcal{M}|$. 
The overall complexity of Algorithm~\ref{alg:OptimumOrchestration} is then $\mathcal{O}\left([T_{m,n} \cdot N_{add} + T_{OBS}] \cdot N_{max} \cdot |\mathcal{M}| \cdot N_{ite} \right)$, where $N_{ite}$ represents the number of iterations of the while loop. This parameter depends on the number of services $|\mathcal{M}|$ as well as the total number of radio resources $N_{cell}^{RB}$. As shown later in Section~\ref{sec:PerformancexApp}, the value of $N_{ite}$ increases with the number of radio resources. Although this is not explicitly reflected in the results for $N_{ite}$, a similar trend has been experimentally observed with respect to the number of services $|\mathcal{M}|$.
\color{black}

\subsection{RT Control Loop}\label{sec:InnerControlLoop}
Every $T_{ OUT}$ \glspl{TTI}, the \gls{RT} Controller \emph{dApp} receives the new value of $N_{ m}^{ min}$ $\forall m \in \mathcal{M}$ from the Delay-Aware \gls{RB} Allocation Controller \emph{xApp} (see point B in Fig.~\ref{fig:ExampleFramework}). Based on them, the \gls{RT} Controller \emph{dApp} operates at each \gls{TTI} as follows. First, it tries to drain the transmission queue of each service $m$ by using $N_{ m}^{ min}$ \glspl{RB}. After, a subset $\mathcal{M}^{ \prime} \subset \mathcal{M}$ of services will have drained their queues, while the remaining $\mathcal{M}^{ \prime\prime}=\mathcal{M}-\mathcal{M}^{ \prime}$  services may still have pending transmissions. Assuming the $|\mathcal{M}^{ \prime}|$ services have not fully consumed their guaranteed amount of \glspl{RB} $N_{m}^{min}$, the total remaining free \glspl{RB} can be represented as $N_{ fr}$. These spare resources, which comprise the sum of the resources not utilized by the $|\mathcal{M}^{ \prime}|$ services, can then be allocated to the remaining $|\mathcal{M}^{ \prime\prime}|$ services (see point D in Fig.~\ref{fig:ExampleFramework}). Specifically, the $N_{ fr}$ \glspl{RB} will be allocated among the $|\mathcal{M}^{ \prime\prime}|$ services following the \gls{EDF} discipline~\cite{EDF_scheduler} since it minimizes the number of packets whose transmission delay is above the target delay budget. Note \gls{EDF} does not consider the violation probability~\cite{Elgabli2019}. For this reason, the \gls{RT} Controller \emph{dApp} combines \gls{EDF} with the establishment of guaranteed \glspl{RB} per service. Since the latter are decided by the Delay-Aware RB Allocation Controller \emph{xApp}, our framework ensures $\mathbb{P}\left[w>W_{ m}\right]\leq \varepsilon_{ m}$ $\forall m \in \mathcal{M}$ as long as the traffic and channel conditions do not change with respect to the samples $\vec{x}_{ a_{ m}}$, $\vec{x}_{ s_{ m},n}$.

If the traffic and/or channel conditions change,  $\mathbb{P}\left[w>W_{ m}\right]$ may be greater than the violation probability $\varepsilon_{ m}$.  To avoid it whenever possible, the \gls{RT} Controller \emph{dApp} executes Algorithm \ref{alg:RT_operation} to temporarily adjust the number of guaranteed \glspl{RB} $N_{m}^{min}$ of the set $\mathcal{M}$ of services (see point E in Fig.~\ref{fig:ExampleFramework}).
\begin{algorithm}[t!]
\SetAlgoLined
\small
\textbf{Inputs:} $N_{ m}^{ min}$, $Q_{ T,m}^{ U}$, $Q_{ T,m}^{ L}$, $\vec{s}_{ i-1}$\;
Compute $q_{ i,m}$ $\forall m \in \mathcal{M}$\;
Update states $\vec{s}_{ i}$ and $N_{ m,i}^{ req}$ according to Fig. \ref{fig:STDtransmissionQueue}\;

$\vec{v}_{ d}=\big\{m^{ \prime}\big\}$ $\forall m^{ \prime}$ $|\vec{s}_{ i}\{m^{ \prime}\}=A$ \;
$\vec{v}_{ b}=\big\{m^{ \prime\prime}\big\}$  $\forall m^{ \prime\prime}$ $|\vec{s}_{ i}\{m^{ \prime\prime}\}=B\cup C$\;
\If{$\vec{v}_d \neq \emptyset$}{
Set $N_{ ite}=\sum_{m^{ \prime\prime}} N_{m^{ \prime\prime},i}^{ req}$ and $N_{ m,i}^{ min} = N_{ m}^{ min}$ $\forall m \in \mathcal{M}$\;
Set $j_{ d} = 0$ and $j_{ b}=0$\;
\For{$u$ from 1 to $N_{ite}$}{
Determine $n^{ \prime} =\vec{v_{ d}}\{j_{ d}\}$ and $n^{ \prime\prime} = \vec{v_{ b}}\{j_{ b}\}$\;
Update $N_{ n^{ \prime},i}^{ min} = N_{ n^{ \prime},i}^{ min} - 1$; $N_{ n^{ \prime\prime},i}^{ min} = N_{ n^{ \prime\prime},i}^{ min} + 1$\;
Update $j_{ d} = j_{ d} + 1$ and $j_{ b} = j_{ b} + 1$\;
\If{$j_{ d} == |\vec{v}_{ d}|$}{
$j_{ d} = 0$
}
\If{$j_{ b} == |\vec{v}_{ b}|$}{
$j_{ b} = 0$
}
}

}
 \textbf{return:} $N_{ m,i}^{ min}$, $N_{ m,i}^{ req}$\;
 \caption{Mitigating $w> W_{ m}^{ th}$ at TTI $i$}
 \label{alg:RT_operation}
\end{algorithm}
\begin{figure}[t!]
    \centering
    \includegraphics[width=\columnwidth]{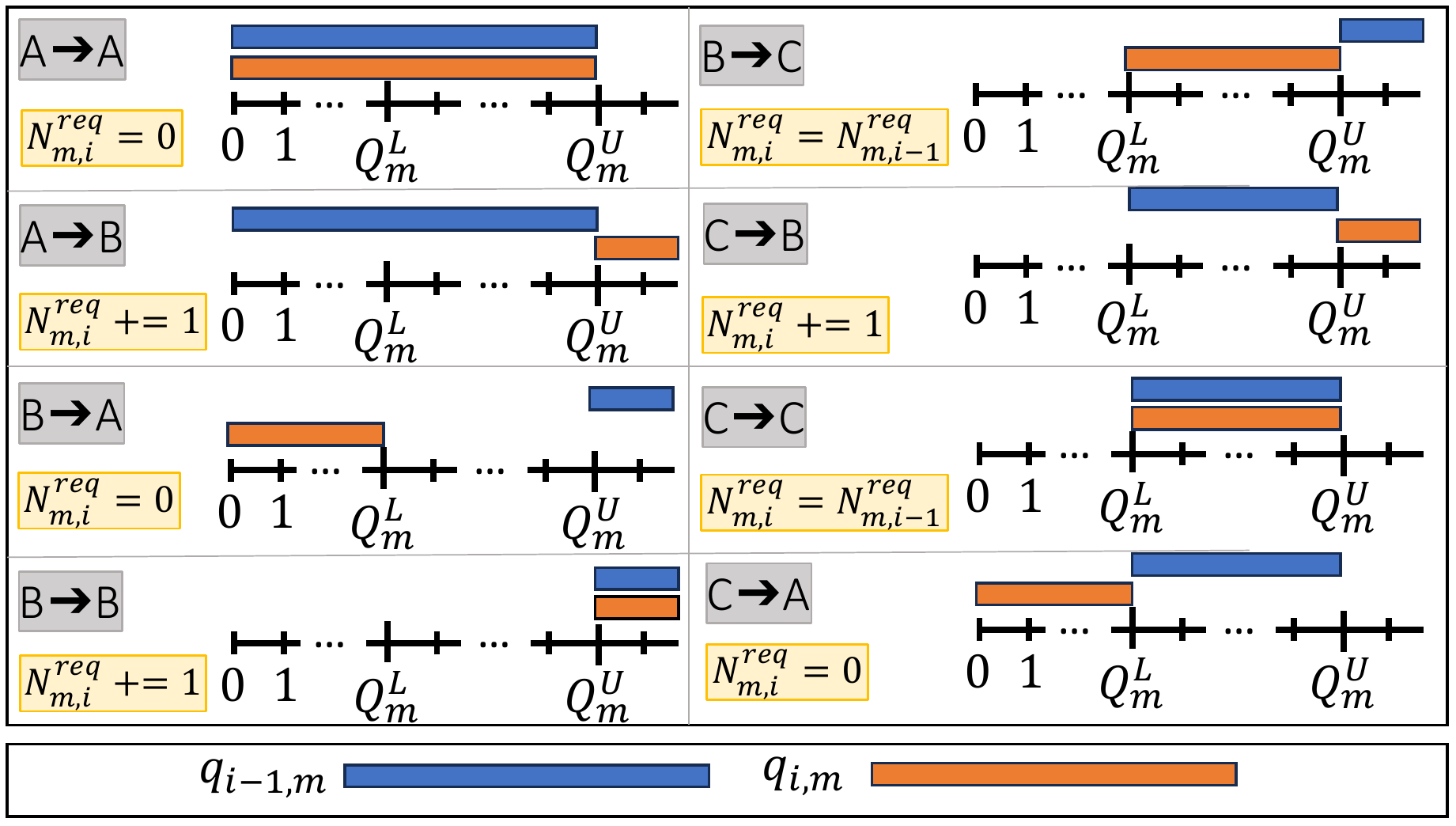}
    \caption{Transitions of the finite-state machine to control the \gls{RB} allocation for service $m$. From \cite{Adamuz2024}.}
    \label{fig:STDtransmissionQueue}
\end{figure}

This algorithm monitors the waiting time of the first packet of each service in the transmission queue. Then, if the waiting time is close to the delay budget, the algorithm increases (if possible) the amount of guaranteed \glspl{RB} for the corresponding service. To this end, Algorithm \ref{alg:RT_operation} relies on a finite-state machine of three states $\big\{A,B,C\big\}$ based on two thresholds $Q_{ T,m}^{ U}$ and $Q_{ T,m}^{ L}$. The threshold $Q_{ T,m}^{ U}=\eta Q_{ T,m}$ indicates the waiting time of a packet is close to the delay budget, whereas $Q_{ T,m}^{ L}= \tau Q_{ T,m}$ indicates the waiting time is far to the delay budget. The parameter $Q_{ T,m}=\lfloor W_{ m}^{ th} / t_{ slot} \rfloor$ is the maximum number of \glspl{TTI} that a packet can wait in the queue before crossing the delay budget. Additionally, $Q_{ T,m}^{ U} > Q_{ T,m}^{ L}$. Note that $\eta \in (0,1]$ and $\tau \in (0,1]$ can be tuned by the \gls{MNO}. Regarding the states, the state $A$ indicates the \gls{RT} Controller \emph{dApp} allocates to the service $m$ the amount of guaranteed \glspl{RB} decided by the Delay-Aware RB Allocation Controller \emph{xApp}, i.e., $N_{ m,i}^{ min} = N_{ m}^{ min}$. Note that we define $N_{ m,i}^{ min}$ as the amount of guaranteed \glspl{RB} decided by the \gls{RT} Controller \emph{dApp} in the \gls{TTI} $i$. The state $B$ indicates the waiting time \textcolor{black}{$w$} of the first packet of service $m$ is very close to the delay budget \textcolor{black}{(i.e., $w/t_{slot}> Q_{T,m}^U$)}, thus the \gls{RT} Controller \emph{dApp} may increase the amount of guaranteed \glspl{RB} for such service. Specifically, it may increase $N_{ m,i}^{ req}$ \glspl{RB}. In state $B$, $N_{m,i}^{ req}$ increases by one \gls{RB} with respect to the previous \gls{TTI}. The state $C$ indicates the waiting time of the first packet is lower than in state $B$, but not enough to go back to  $N_{ m,i}^{ min} = N_{ m}^{ min}$. In such case, Algorithm \ref{alg:RT_operation} keeps the same value of $N_{ m,i}^{ req}$ with respect to the previous \gls{TTI}. In Fig.~\ref{fig:STDtransmissionQueue} we summarize the possible transitions among states. Considering such transitions, we define $\vec{s}_{ i}$ as a vector containing the state for each service at \gls{TTI} $i$.

Based on $N_{ m}^{ min}$, $Q_{ T,m}^{ L}$, $Q_{ T,m}^{ U}$ and $\vec{s}_{ i-1}$, Algorithm \ref{alg:RT_operation} initially computes the state of the first packet of each service as $q_{ i,m}=w_{ m}^{ pkt}/t_{ slot}$ (step 2). Note that $w_{ m}^{ pkt}$ is the waiting time of such a packet. Then, it updates the states $\vec{s}_{ i}$ and $N_{ m,i}^{ req}$ according to the transitions depicted in Fig. \ref{fig:STDtransmissionQueue} (step 3). Later,  Algorithm \ref{alg:RT_operation} needs to check if the amount of \glspl{RB} defined in $N_{ m,i}^{ req}$ can be allocated, in addition to $N_{ m}^{ min}$, to the corresponding services. The policy considered by Algorithm~\ref{alg:RT_operation} is that only the services whose state is $A$ can donate \glspl{RB} to those which require more \glspl{RB}. Considering this policy, Algorithm \ref{alg:RT_operation} iteratively re-allocates the amount of guarantees \glspl{RB} from services in state $A$ to services in states $B$ or $C$ (steps 4-20).  

\color{black}
The computational complexity of Algorithm~\ref{alg:RT_operation} can be generically expressed as $\mathcal{O}(|\mathcal{M}| + N_{ite})$. In particular, lines 2-5 involve checking the transmission queue state for each service $m\in\mathcal{M}$, while the remaining lines depend on the number of iterations $N_{ite}$. In turn, $N_{ite}$ is governed by two key factors. First, there must exist services capable of donating radio resources (see line 4 in Algorithm~\ref{alg:RT_operation}), meaning their transmission queues must be in state A. If this condition is met, then $N_{ite}$ depends on the number of additional resources requested by the services whose transmission queues are in state B or C (see line 7 in Algorithm~\ref{alg:RT_operation}). Both $|\mathcal{M}|$ and $N_{ite}$ depend on the number of services as well as the state of their transmission queues at the time Algorithm 3 is executed. 

\color{black}

\section{Performance Results}\label{sec:PerformanceResults}
\begin{figure}[t!]
    \centering
    \includegraphics[width=\columnwidth]{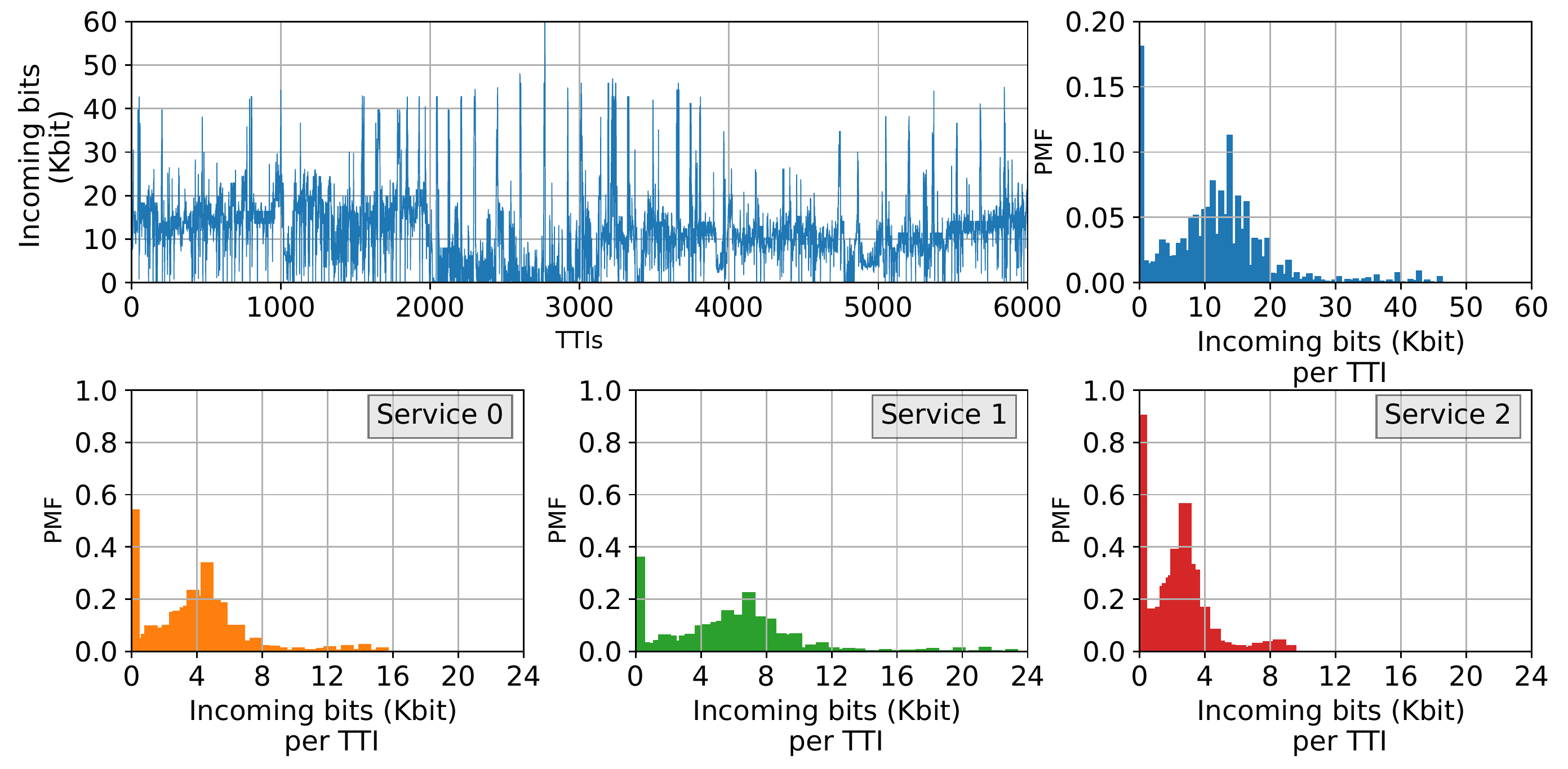}
   \caption{PMFs of incoming bits from FALCON traces. Copied from \cite{Adamuz2024}.}
    \label{fig:InputServices}
\end{figure}

We conducted an extensive simulation campaign to validate \name{} and assess its performance using a Python-based simulator running on a computing platform with $16$ GB RAM and a quad-core Intel Core i7-7700HQ @ 2.80 GHz. The simulator models a single cell utilizing an \gls{OFDMA} scheme with $N_{ cell}^{ RB}\in [50,100]$ \glspl{RB} and $t_{ slot}$ = 1 ms. To simulate the traffic and cell capacity of each uRLLC service, we use realistic traces collected from an operational RAN with the FALCON tool~\cite{falcon}.
FALCON enables the decoding of the Physical Downlink Control Channel (PDCCH) of a \gls{vBS}, providing insight into the number of active users and their allocated resources.
Fig.~\ref{fig:InputServices} illustrates the incoming bits per \gls{TTI} measured by FALCON, along with the corresponding \gls{PMF} in the upper plots. To emulate the incoming traffic for three distinct \gls{uRLLC} services, we sorted the active \glspl{UE} and divided them into equally sized groups based on their aggregated traffic demand. The resulting \glspl{PMF} for these groups are also shown in the lower plots of Fig.~\ref{fig:InputServices}.
For these services, we set a delay budget $W_{ m}^{ th}=\big\{5,10,15\big\}$ ms and target violation probability $\varepsilon_{ m}=\big\{10^{-5},10^{-4},10^{-3}\big\}$ \cite{Haque2023,Siddiqui2023,Khan2022}.

\subsection{Validation Martingales-based Model}\label{sec:MartingalesValidation}

In the first experiment, we examine a cell hosting a single service $m$ with incoming traffic measured by FALCON (i.e., blue plots in Fig.~\ref{fig:InputServices}). Assuming a target violation probability $ \varepsilon_{ m}=10^{-3} $, we estimate the delay bound using the proposed Martingales-based model and compare it with the real delay bound obtained through simulation. We also compare this with delay bounds estimated using the \gls{SNC}-based model proposed in \cite{Adamuz2024}. The Delay-Aware RB Allocation Controller \emph{xApp} (a) allocates $ N_{ m}^{ min} \in \left[40,100\right] $ \glspl{RB} for the service, and (b) uses $ T_{ OBS} \in \big\{1, 2, 3, 4, 5, 6 \big\} \cdot 1000 $ \glspl{TTI} to obtain the sample vectors $ \vec{x}_{ a_m} $ and $ \vec{x}_{ s_{ m},n} $.

\begin{figure}[t!]
    \centering
    \includegraphics[width=\columnwidth]{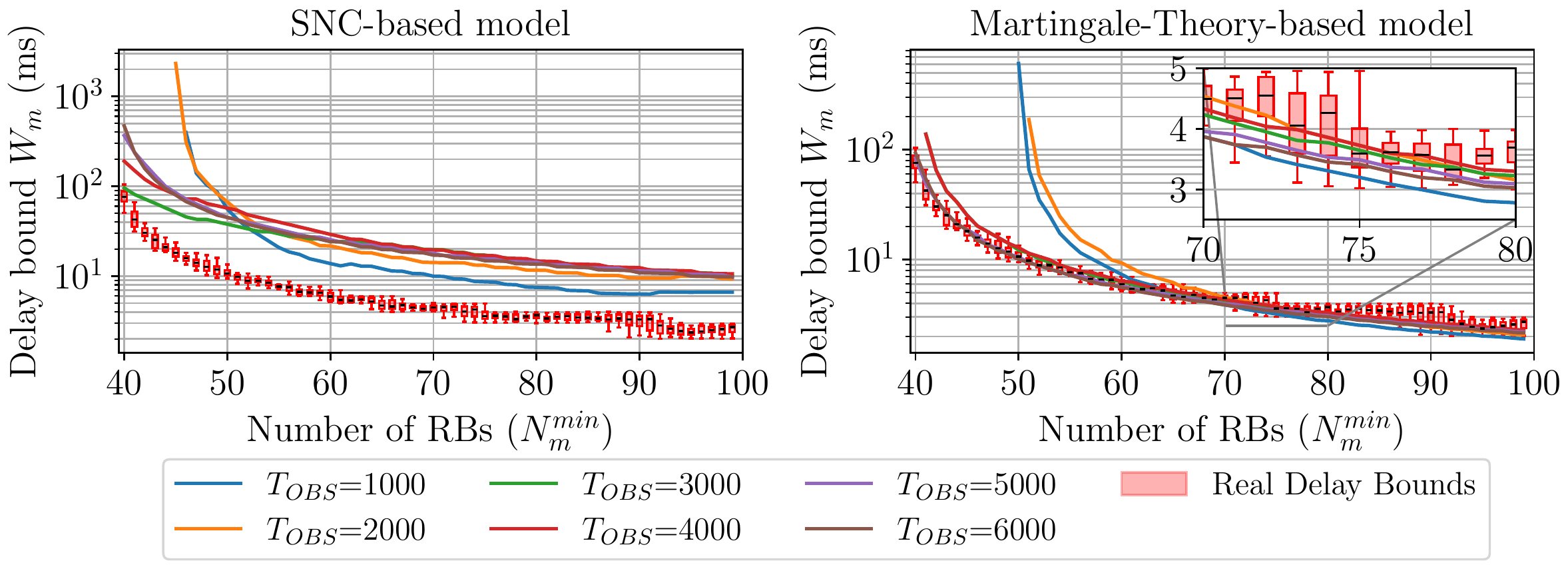}
    \caption{Evaluation of delay bound $W_m$ for a service $m$ based on the number of guaranteed \glspl{RB} $N_m^{min}$ and considering a specific metric collection period size $T_{OBS}$.}
    \label{fig:ValidationDelayBounds}
\end{figure}

Fig.~\ref{fig:ValidationDelayBounds} illustrates the evolution of the delay bound as the Delay-Aware RB Allocation Controller \emph{xApp} allocates a specific number of \glspl{RB} to the service. The left plot depicts the delay bound estimation using the \gls{SNC}-based model, while the right plot represents the estimation from the proposed Martingales-based model. Both plots include red boxplots representing the delay bounds measured experimentally through simulation. For each specific \gls{RB} allocation, the delay bound was experimentally measured 50 times, with each measurement based on a simulation of 4 million \glspl{TTI}.

\begin{figure}[t!]
    \centering
    \includegraphics[width=\columnwidth]{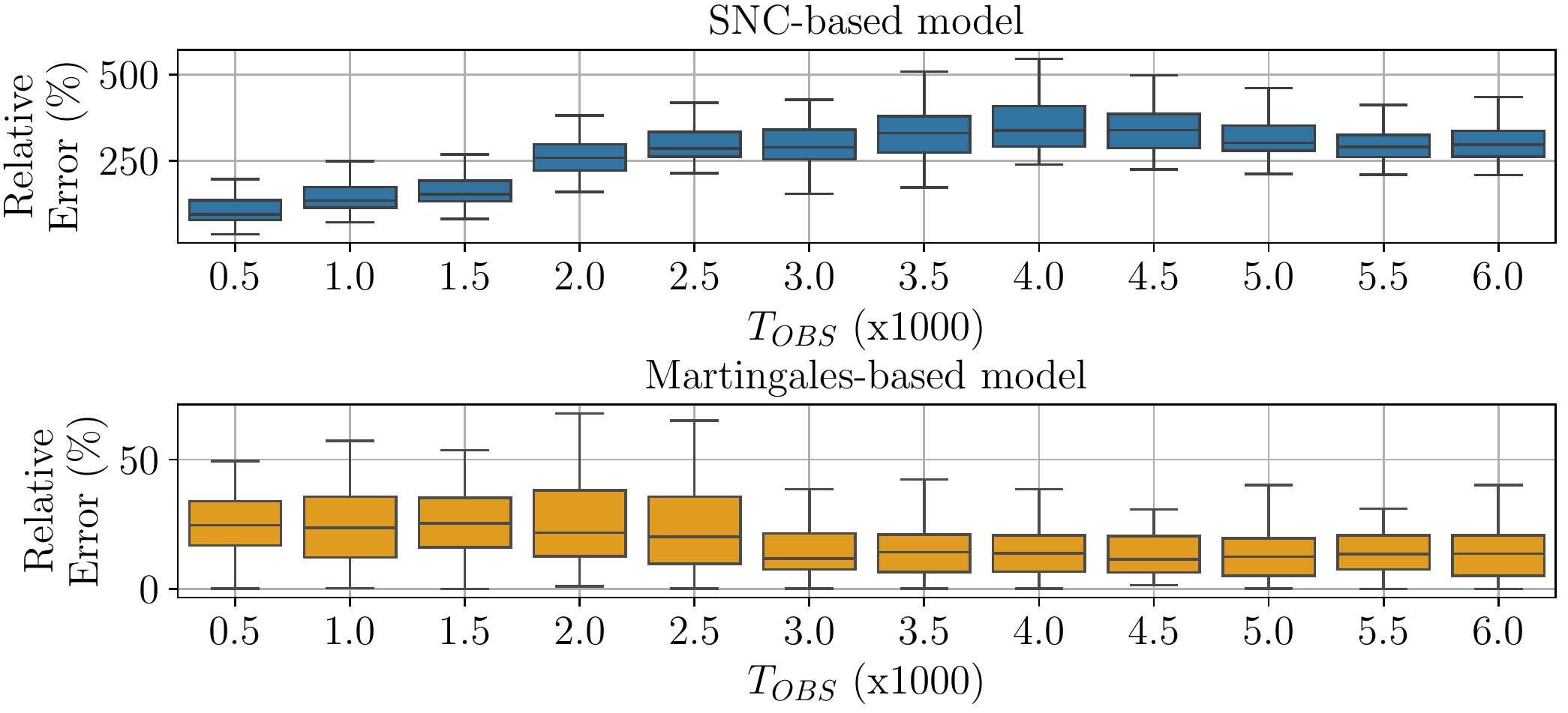}
    \caption{Evaluation of the relative error between the estimated delay bound and the delay bound measured through simulation, considering a specific metric collection period size $T_{OBS}$. }
    \label{fig:RelativeErrorDelayBounds}
\end{figure}

The results show that delay bounds from the \gls{SNC} model are significantly less accurate than those from the proposed Martingales-based model, which closely aligns with real values. To quantify this difference, Fig.~\ref{fig:RelativeErrorDelayBounds} illustrates the relative error between the actual delay bounds (measured in simulations) and those estimated by both models. The average relative error ranges from 15\% to 25\% for the Martingales-based model, compared to around 300\% for the \gls{SNC} model, demonstrating superior accuracy.

When comparing measurements for different $T_{OBS}$ values, both models provide reliable estimates when $T_{OBS} \geq 4000$. Below this threshold, estimations are less accurate, especially with fewer \glspl{RB} are allocated. For the Martingales-based model, we can see that the delay bound estimation is significantly higher than the real value for the blue and orange curves ($T_{OBS}=1000$ and $T_{OBS}=2000$, respectively).
For the \gls{SNC} model, even with the green curve ($T_{OBS}=3000$), the estimation is no longer conservative (i.e., the estimated value exceeds the real value). Note that the \gls{SNC}-based model guarantees accurate upper-bound estimations only when the \gls{PMF} for both arrival and service processes is correctly captured~\cite{Adamuz2024}. Due to insufficient samples when $T_{OBS} < 4000$, this is not the case. This fact also explains why the \gls{SNC} model shows a "lower" relative error with few samples. However, this lower error is misleading; the estimation is not better but appears so because, with insufficient samples, the estimates and actual values intersect at lower RB values, creating a false impression of improved accuracy.

\begin{figure}[t!]
    \centering
    \includegraphics[width=\columnwidth]{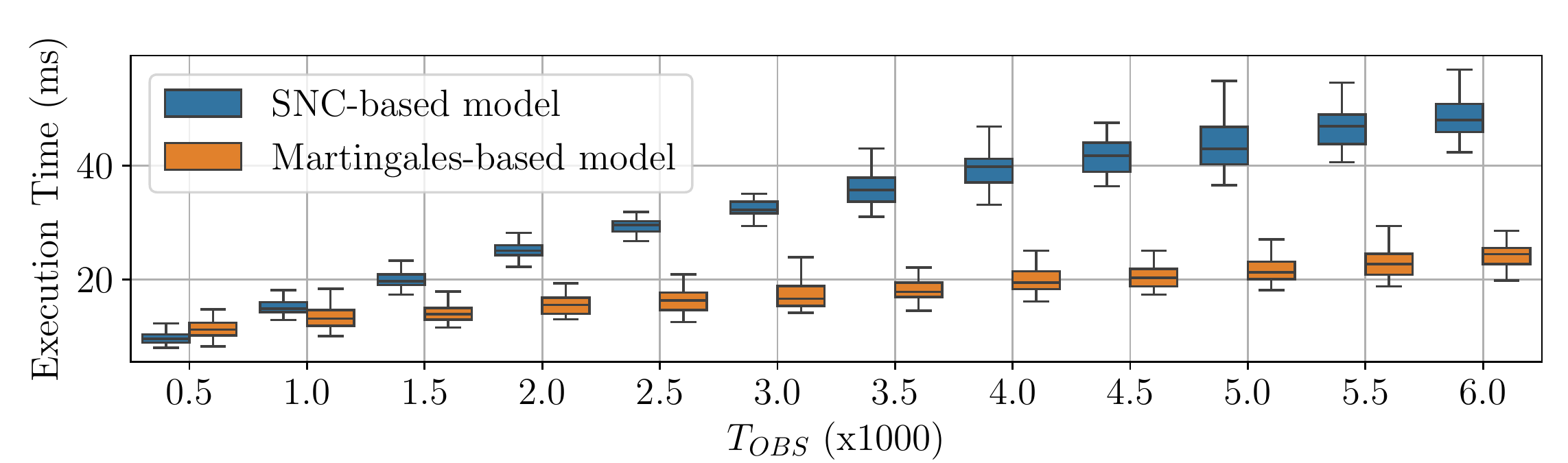}
    \caption{Execution time comparison between the Martingales-based model and the \gls{SNC}-based model for estimating the delay bound $W_m$.}
    \label{fig:ExecutionTimeMartingales}
\end{figure}

The execution times of the Martingales-based model and the \gls{SNC}-based model were also compared, as shown in Fig.~\ref{fig:ExecutionTimeMartingales}. The boxplot illustrates that both models exhibit a linear increase in execution time with $T_{OBS}$, but the \gls{SNC}-based model's time grows more rapidly. For $T_{OBS}=6000$, the \gls{SNC}-based model's execution time is twice that of the Martingales-based model. This is due to the Martingales-based model only needing to optimize one parameter, $\theta_{m}^{\ast}$, while the \gls{SNC}-based model optimizes two parameters~\cite{Adamuz2024}.

Analyses of the previous results show that the Martingales-based model provides better delay bound estimations and faster execution times than the \gls{SNC}-based model, making it ideal for integration into the proposed framework.

\subsection{Performance Analysis of the Delay-Aware RB Allocation Controller xApp}\label{sec:PerformancexApp} 
In a third set of experiments, we focus on a single decision period of the Delay-Aware RB Allocation Controller \emph{xApp} when it consider both, the Martingales-based model and the \gls{SNC}-based model~\cite{Adamuz2024} to estimate the delay bound. Specifically, we assume this \emph{xApp} computes $N_{ m}^{ min}$ for the three services described at the beginning of Section~\ref{sec:PerformanceResults}. Under this scenario, we evaluate the convergence of the heuristics proposed in Algorithm \ref{alg:OptimumOrchestration}, the computational complexity of such heuristics, and the accuracy of the obtained solution with respect to the optimal.
\begin{figure}[t!]
    \centering
    \includegraphics[width=\columnwidth]{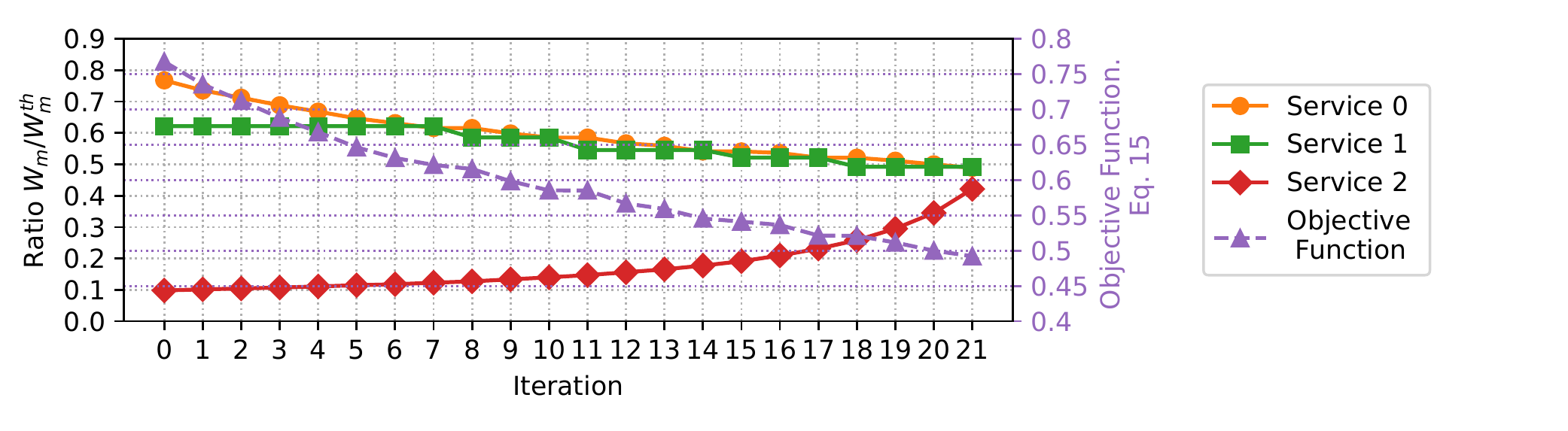}
    \caption{Convergence Analysis of Algorithm 
    \ref{alg:OptimumOrchestration}.}
    \label{fig:Orchestrator}
\end{figure}

First, we evaluate the convergence of Algorithm \ref{alg:OptimumOrchestration}. We consider the Delay-Aware RB Allocation Controller \emph{xApp} applies the Martingales-based model to derive the delay bound. Fig.~\ref{fig:Orchestrator} depicts the objective function value $g(\vec{W})$, i.e., purple curve, and the ratios $W_{ m}/W_{ m}^{ th}$ $\forall m \in \mathcal{M}$, i.e., orange, green and red curves, across iterations. The heuristics progressively reduce $g(\vec{W})$ until achieving the optimal solution.

\begin{table}[t!]
\centering
\caption{Performance Comparison of Delay-Aware RB Allocation Controller xApp using SNC-based and Martingales-based models with a brute force algorithm.}
\label{tab:BruteForceComparison}
\resizebox{\columnwidth}{!}{
\begin{tabular}{|c|c|c|c|c|c|}
\hline
\rule{0pt}{1.2em} 
\textbf{$\mathbf{N_{cell}^{RB}}$}                                             & \textbf{60} & \textbf{70} & \textbf{80} & \textbf{90} & \textbf{100} \\[4pt] \hline
\textbf{\begin{tabular}[c]{@{}c@{}} Relative error (\%)\\ 
SNC-based model\end{tabular}}  & 23.07        & 21.43            & 23.05        & 21.43          & 17.35           \\ \hline
\textbf{\begin{tabular}[c]{@{}c@{}} Relative error (\%)\\ 
Maringale-based model\end{tabular}}  & 0        & 0            & 0        & 0          & 0           \\ \hline
\textbf{\begin{tabular}[c]{@{}c@{}}Brute Force (iterations)\end{tabular}}  & 1711        & 2346           & 3081        & 3916        & 4851         \\ \hline
\textbf{\begin{tabular}[c]{@{}c@{}}Algorithm 2 (iterations) \\ SNC-based model\end{tabular}} & 12           & 14          & 16                  & 19                & 22           \\ \hline
\textbf{\begin{tabular}[c]{@{}c@{}}Algorithm 2 (iterations) \\ Martingales-based model\end{tabular}} & 12           & 14          & 16                  & 20                & 22           \\ \hline
\textbf{\begin{tabular}[c]{@{}c@{}} Ratio iterations \\ SNC-based model\end{tabular}} & 142.58           & 167.57          & 192.56                  & 206.10               & 220.5           \\ \hline
\textbf{\begin{tabular}[c]{@{}c@{}} Ratio iterations \\ Martingales-based model\end{tabular}} & 142.58           & 167.57          & 192.56                  & 195.8                & 220.5           \\ \hline
\textbf{\begin{tabular}[c]{@{}c@{}} Avg. Execution Time per \\ iteration (ms) when \\ SNC-based model\end{tabular}} & 184.62            & 179.09             & 176.39                & 173.76                & 171.85            \\ \hline
\textbf{\begin{tabular}[c]{@{}c@{}} Avg. Execution Time per \\ iteration (ms) when \\ Martingales-based model\end{tabular}} & 106.38              & 99.67             & 91.63                & 88.95                & 87.90            \\ \hline
\end{tabular}
}
\end{table}

Second, we compare the heuristic solution with the optimal one obtained via brute force for $N_{cell}^{RB} \in [60,100]$ RBs. The heuristics employ both the Martingales-based model and the \gls{SNC}-based model~\cite{Adamuz2024}. Table~\ref{tab:BruteForceComparison} shows that the heuristic with the Martingales-based model reaches the optimal solution, while the SNC-based model yields a near-optimal solution with a 20\% relative error. This confirms that our heuristic can provide near-optimal solutions efficiently, avoiding the computationally expensive brute force method.

Furthermore, we observe the number of iterations increases with $N_{cell}^{RB}$ due to the larger search space, while the average execution time per iteration decreases slightly. This is because estimating $W_{m}$ (via Algorithm~\ref{alg:Search-for-optimum-theta} to determine $\theta_m^{\ast}$ and $W_{m}$) becomes faster with more \glspl{RB}. Specifically, fewer iterations are needed as $N_{cell}^{RB}$ increases, as previously shown in Fig~\ref{fig:ValidationKaKs}. Comparing execution times, the Delay-Aware RB Allocation Controller \emph{xApp} performs iterations faster with the Martingales-based model than with the \gls{SNC}-based model, as the former estimates delay bounds more quickly as previously depicted in Fig.~\ref{fig:ExecutionTimeMartingales}.

\subsection{Service Accommodation Capability of the Delay-Aware RB Allocation Controller xApp}

\begin{figure}[b!]
    \centering
    \includegraphics[width=\columnwidth]{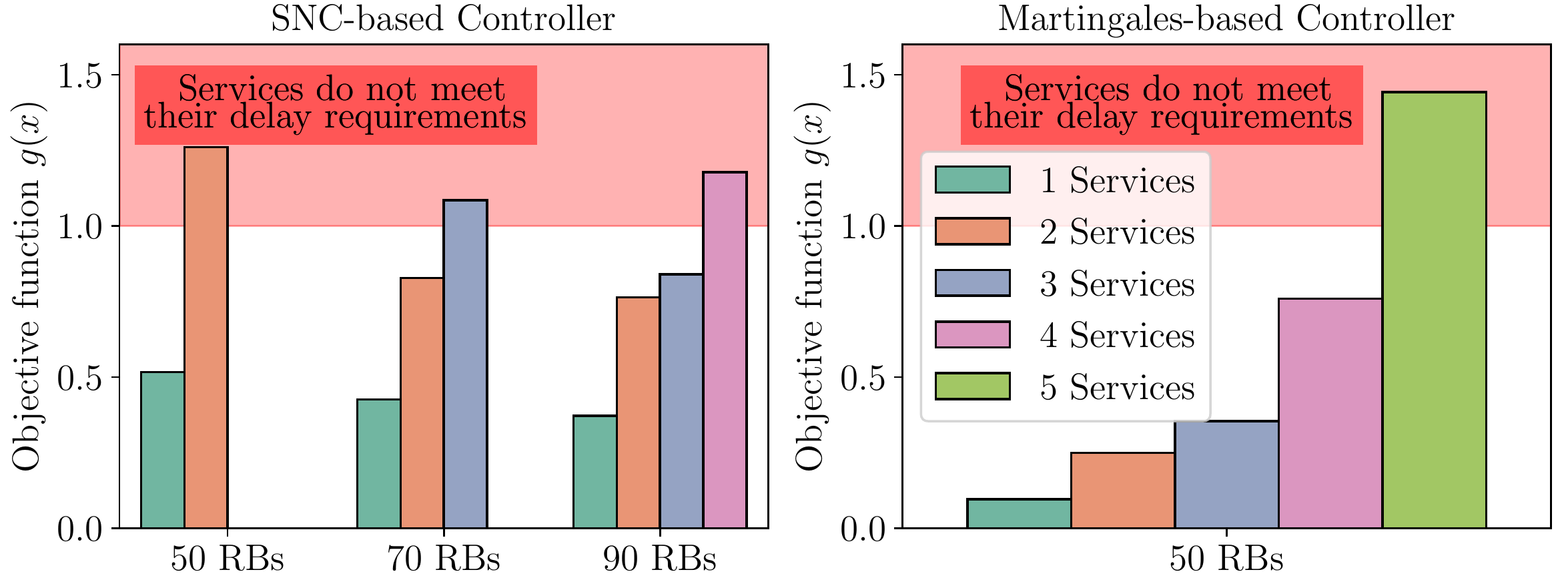}
    \caption{Analysis of the service accommodation capability when the cell has available a specific amount of \glspl{RB}.}
    \label{fig:ServiceAccomodation}
\end{figure}

In this experiment, we evaluated the service accommodation capability of the Delay-Aware \gls{RB} Allocation Controller \emph{xApp}, which can use either the proposed Martingales-based model or the \gls{SNC}-based model  presented in~\cite{Adamuz2024}. Specifically, we assessed each controller's ability to allocate radio resources to multiple coexisting \gls{uRLLC} services within a cell, ensuring that the delay requirements of these services are met simultaneously. This means the objective function $g(\vec{W})$ defined in Eq.~\eqref{eq:OptProblem} must be less than or equal to 1.

Results are presented in Fig.~\ref{fig:ServiceAccomodation}, which shows the value of the objective function  $g(\vec{W})$ when deploying a specific number of \gls{uRLLC} services given a fixed number of \glspl{RB} available in the cell. The left plot displays the objective function value for the \gls{SNC}-based controller, while the right plot shows the value for the Martingales-based controller. Note that both controllers were evaluated under the same conditions, including identical services with the same latency requirements, traffic generation distribution, and channel conditions.

The results indicate that the Martingales-based controller can accommodate up to 4 services simultaneously with 50 \glspl{RB}, whereas the \gls{SNC}-based controller can only accommodate a single service with the same number of \glspl{RB}. Even with 90 \glspl{RB}, the \gls{SNC}-based controller can handle only 3 services, which is fewer than the number of services accommodated by the Martingales-based controller with 50 \glspl{RB}. Although services orchestrated by the \gls{SNC}-based controller would experience lower delay bounds (see results of Section \ref{sec:MartingalesValidation}), the conservative estimations of the \gls{SNC}-based model limits the number of services that can be deployed by the \gls{MNO}. In contrast, the Martingales-based model provides more accurate delay bound estimations, allowing the proposed controller to deploy a greater number of services. This highlights the significant advantage of using a Martingales-based model over an \gls{SNC}-based model for orchestrating multiple \gls{uRLLC} services.

\subsection{Performance Analysis of \name{} Framework}
In the last experiment, we evaluated the performance of \name{} against three reference solutions. \textcolor{black}{These solutions are based on different resource allocation mechanisms and control strategies, as detailed below:} 
\begin{itemize}
    \item \textbf{Reference solution \#1}: only considers a single \gls{RT}-Controller (\textit{dApp}) in the \gls{O-RAN} architecture, located in the \gls{CU}-\gls{CP}. The RT-Controller implements an \gls{EDF} scheduler across all $|\mathcal{M}|$ services. For each service $m\in\mathcal{M}$, a transmission queue is considered, and the scheduler monitors the waiting time of the first packet in each queue. The packet whose waiting time is closest to its corresponding delay bound $W_{m}^{th}$ is transmitted first.
    \item \textbf{Reference solution \#2}: the Queue Length and Delay Requirement (QLDR)-Based Algorithm, proposed in \cite{Dai2025}, operates in two main steps. First, for each service $m\in\mathcal{M}$, the average transmission queue length (in bits) and the average spectral efficiency (in bits/s/Hz) are computed over the past $T$ \glspl{TTI}. Then, the average queue length is normalized by the average spectral efficiency and further divided by the service's target delay bound $W_{m}^{th}$. Finally, the available \glspl{RB} $N_{RB}^{cell}$ are allocated among these services in proportion to the previous normalized values. Note that $N_m^{min}$  $\forall m\in \mathcal{M}$ is updated every $T$ \glspl{TTI}.
    \item \textbf{Reference solution \#3}: utilizes all the \textit{xApps} proposed in \name{} framework, except for the \gls{RT} Controller \textit{dApp}. Specifically, the Delay-Aware \gls{RB} Allocation Controller \textit{xApp} is responsible for periodically calculating the \glspl{RB} required by each service $m\in\mathcal{M}$, as in \name{}. However, a key distinction is that this solution does not allow for \gls{RB} sharing between services during the assignment period. Each service $m\in\mathcal{M}$ is allocated dedicated $N_m^{min}$ \glspl{RB}, and those resources that are not being used by a service (in specific \glspl{TTI}) cannot be shared with other services, unlike \name{}, which allows sharing of unused \glspl{RB} between services within the same assignment period.
    \item \textbf{Reference solution \#4}: the \gls{RT} Controller \textit{dApp} is included, along with the remaining \name{}'s \textit{xApps}, enabling the sharing of \glspl{RB} among services. Specifically, if certain services are not utilizing their full allocated \glspl{RB} at specific \glspl{TTI}, those unused \glspl{RB} can be reassigned to other services with higher instantaneous traffic load. However, the \gls{RB} sharing is implemented using an \gls{EDF} scheduler, without considering Algorithm~\ref{alg:RT_operation}, which is a key aspect of \name{}. In our framework, the \gls{EDF} scheduler is complemented by Algorithm~\ref{alg:RT_operation}, which takes into account the state of the transmission queues for each service $m\in\mathcal{M}$, providing a more sophisticated approach to resource sharing. 
\end{itemize}
\color{black}


\begin{figure}[t!]
    \centering
    \includegraphics[width=\columnwidth]{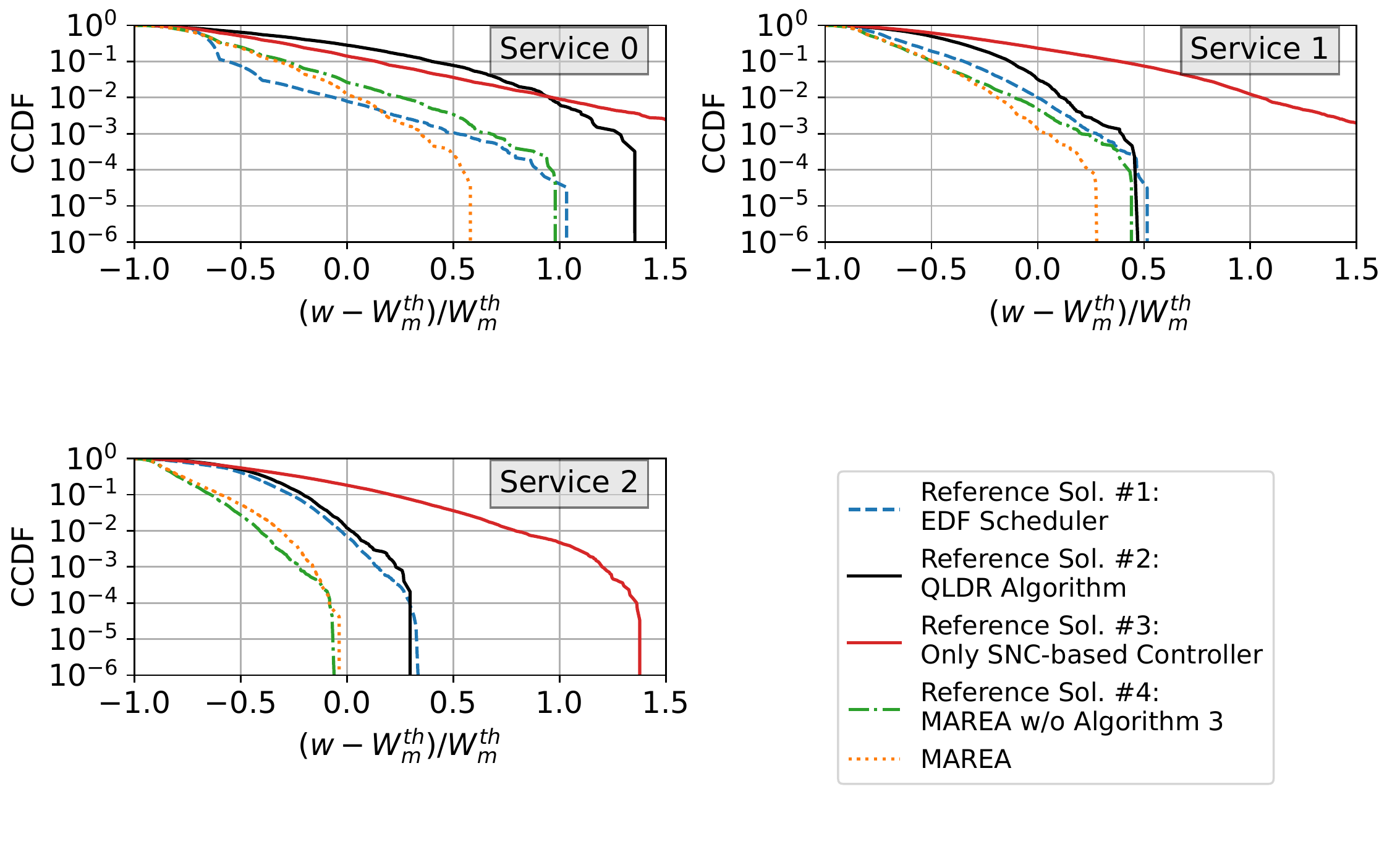}
   \caption{Complementary Cumulative Distribution Function (CCDF) for \\ $(w-W_{ m}^{ th})/W_{ m}^{ th}$. In \name{}, we have set  $\eta = 0.75$ and $\tau = 0.3$.}
   \label{fig:SLA_analysis}
\end{figure}

To measure the performance of these solutions, we consider the \gls{CCDF} of the metric $(w-W_{ m}^{ th})/W_{ m}^{ th}$. Note the random variable $w$ represents the transmission delay of an arbitrary packet. Additionally, when this metric is equal to 0, the \gls{CCDF} value represents the violation probability. In Fig.~\ref{fig:SLA_analysis} we observe that the reference solution \textcolor{black}{\#3} provides the worst performance, i.e., a violation probability $11.46$ and $178.30$ times larger than \name{} for services 1 and 2. Note that this probability is equal to 0 for scenario 3 when we consider \name{}. These results are due to the fact the reference solution \#2 only considers dedicated \glspl{RB}. \textcolor{black}{Reference solution \#2 also considers dedicated \glspl{RB}, but outperforms reference solution \#3. This is because its dedicated \gls{RB} allocation is updated every 10 \glspl{TTI}, whereas in reference solution \#3, it is updated every 1000 \glspl{TTI}. As a result, despite not using a delay estimation model, reference solution \#2 better accounts for the current transmission queue state, capturing the impact of unexpected traffic behavior on queue occupancy and the current radio channel capacity for these users.} The remaining solutions consider sharing \glspl{RB} among the services. Comparing them, the reference solution \#1 usually provides a greater violation probability with respect to \name{}. Although \gls{EDF} ensures the packet with the earliest deadline are transmitted first, e.g., we observe for service 0 the \gls{CCDF} is lower for reference solution \#1 when $(w-W_{ m}^{ th})/W_{ m}^{ th} \leq 0$, it does not provide any guarantees in terms of violation probability. To consider such probability, the \name{} framework establishes guaranteed \glspl{RB} in a near-\gls{RT} and allocates share free \glspl{RB} among services using \gls{EDF} in a \gls{RT} scale. It improves the performance with respect to the remaining solutions. Specifically, \name{} provides lower violation probabilities for service \#1 and service \#2. Finally, we observe the consideration of Algorithm \ref{fig:STDtransmissionQueue} in \name{} improves the obtained violation probability if we compare the results with the ones obtained by the reference solution \textcolor{black}{\#4}. The improvement is most significant when the metric $(w-W_{ m}^{ th})/W_{ m}^{ th}$ is higher. This is due to Algorithm \ref{fig:STDtransmissionQueue} performing its actions when the waiting time of a packet in the transmission queue is closer to the delay budget. We can also observe the reference solution \textcolor{black}{\#4} has a similar behavior as \name{} for service 2. The packet transmission delay is never above the delay budget for this service, as happens for services 0 and 1.

\section{Challenges for Implementing \name{} in Real-world O-RAN Deployments}\label{sec:Challenges}
The deployment of \name{} in an \gls{O-RAN} testbed would allow to evaluate its performance in real radio conditions based on specific deployment scenarios. However, implementing \name{} in such an environment presents several challenges that require extensive investigation and are thus beyond the scope of this work.

\textbf{Real-Time Task Execution.} A key challenge is ensuring real-time task execution across the different functional blocks of \name{}. Its two control loops, working at near-\gls{RT} and \gls{RT} timescales, need fast processing to make sure radio resource allocations happen on time. In particular, the \gls{RT} control loop needs hardware optimization to adjust radio resource allocation at each \gls{TTI} or every few \glspl{TTI}, depending on the transmission queue status of \gls{uRLLC} services. This can be done using hardware accelerators like FPGAs and/or GPUs to run these time-critical tasks efficiently. The advantages of such hardware acceleration in \gls{O-RAN} systems have been shown in testbeds like X5G~\cite{Villa2024,Villa2024-2}, which uses NVIDIA Aerial to offload physical layer tasks, improve performance through parallel processing, and enable AI/ML-based optimization.

\textbf{Compatibility with standardized interfaces.} To integrate \name{}’s \emph{xApps} and \emph{dApp} with existing \gls{O-RAN} components, their functionalities need to be adapted so that their input and output data match the \gls{O-RAN} standardized APIs. This involves adjusting data formats and ensuring they can interact with the \texttt{E2} interface for real-time radio resource control~\cite{o-ran1,o-ran2,o-ran3,o-ran4}. 

\textbf{Access to Licensed Spectrum.} Access to licensed spectrum is one of the key challenges in integrating \name{} into a commercial \gls{O-RAN} platform. Obtaining licenses for \gls{5G} frequency bands is often complicated due to regulatory restrictions. However, in some countries, certain frequency bands, such as those in the 3.5 GHz range, have been partially allocated for industrial and research use, offering more flexibility for testing~\cite{pedersen2024}. These frequency allocations, available through regulatory sandboxes or collaborative testbed initiatives~\cite{Upadhyaya2023}, could enable more adaptable experimentation. 

\section{Related Works}\label{sec:RelatedWorks}
Several studies have explored radio resource allocation in networks where \gls{eMBB} and \gls{uRLLC} services coexist, leveraging puncturing to minimize \gls{uRLLC} packet transmission delay. Bairagi et al.\cite{Bairagi2021} formulate an optimization problem to maximize the minimum expected \gls{eMBB} rate while ensuring efficient \gls{uRLLC} allocation. Their approach combines Penalty Successive Upper Bound Minimization (PSUM) for \gls{eMBB} scheduling with a Transportation Model (TM) for \gls{uRLLC}, improving fairness and minimum achieved rates. Similarly, Alsenwi et al.\cite{Alsenwi2019} propose a risk-sensitive slicing framework that optimizes \gls{eMBB} puncturing probability while maintaining reliability. Their method models the tail distribution of \gls{eMBB} rates and employs convex relaxation for iterative optimization. For a comprehensive review of \gls{uRLLC} puncturing, see~\cite{Haque2023}. However, these works do not address networks with only \gls{uRLLC} services or integrate O-RAN architectural constraints.

Other authors have addressed radio resource allocation in networks with multiple \gls{uRLLC} services. In the context of \gls{O-RAN}, Abedin et al.\cite{Abedin2022} propose an actor-critic framework to minimize the probability of \gls{IoT} devices exceeding an age of information threshold, though it overlooks air interface transmission delay. Karbalaee et al.\cite{Karbalaee2023} introduce an iterative algorithm for joint radio resource and power allocation, while Rezazadeh et al.\cite{Rezazadeh2023} address the problem via federated learning—both focusing primarily on average delay. Polese et al.\cite{PoleseColo} evaluate \gls{DRL} agents within a non-\gls{RT} control loop in \gls{O-RAN}. Despite their contributions, these works lack details on multi-time-scale control loop interactions.

Non-\gls{RT} and near-\gls{RT} solutions often rely on queuing theory to model packet transmission delay~\cite{Zhang2023,Shi2022,Yang2021}, but these models only yield average values under complex arrival and channel capacity distributions. Alternatively, some works~\cite{SOTANC1,SOTANC2,SOTANC3} use \gls{SNC} to estimate a delay bound $W$ of type $\mathbb{P}[w>W]<\varepsilon$, where $w$ is the delay and $\varepsilon$ a target tolerance. However, they do not account for multiple \gls{uRLLC} services or cross-interference. In~\cite{Adamuz-Hinojosa-TWC2023}, we proposed an \gls{SNC}-based controller for planning multiple \gls{uRLLC} services, but it assumes dedicated resources, potentially leading to inefficiencies, and is limited to Poisson batch arrivals. Recently, we introduced an \gls{O-RAN}-compliant framework based on a novel \gls{SNC} model for multi-service \gls{uRLLC} allocation~\cite{Adamuz2024}, incorporating real traffic and capacity metrics. However, \gls{SNC}-based delay bounds often deviate significantly from actual values, restricting the number of deployable services. Martingale queueing methods have demonstrated significantly improved delay bound estimations. Poloczek and Ciucu derive tight stochastic delay bounds for Markovian sources over ALOHA and CSMA/CA~\cite{Ciucu2014,Poloczek2015}, while Zhao et al. optimize delay and energy efficiency in \gls{mMTC} using differentiated ALOHA~\cite{Zhao2018}. Other works analyze end-to-end delay in multi-hop networks~\cite{Fantacci2021} and multiplexing of \gls{uRLLC} and \gls{eMBB} traffic using reconfigurable intelligent surfaces~\cite{Peng2024}. Yu et al. propose frameworks for bandwidth abstraction and network reliability under strict latency constraints~\cite{Yu2020}. Despite their advancements, these studies assume well-known traffic distributions (Poisson, Bernoulli, or Markov on-off), limiting their applicability to real-world traffic patterns with arbitrary distributions.


Considering \gls{RT} solutions, ~\cite{EDF_scheduler,Hadar2018,Raviv2023} use schedulers based on \gls{EDF} to assign priorities to packets based on their deadlines. \gls{EDF} ensures the packets with the earliest deadline are transmitted first. However, \gls{EDF} does not consider the probability the packet transmission delay exceeds a delay budget~\cite{Capozzi2013}. Other solutions such as~\cite{Zhang2019,Esswie2020,Alsenwi2021} rely on \gls{ML} models. Although they are effective in managing scenarios with intricate traffic patterns and channel conditions, their performance is primarily reliant on the similarity between the measured patterns and those used during training.

\gls{O-RAN} specifications mention a \gls{RT} control loop for optimizing tasks such as packet scheduling or interference recognition~\cite{O-RAN-WG2-AIML}. However, at the moment of writing this paper, such a control loop has not been defined. In the same row, the authors of \cite{dApps_article} introduce the concept of \emph{dApps} to implement fine-grained \gls{RT} control tasks. Despite implementing a proof-of-concept, they omit to detail how multiple \gls{uRLLC} services can be orchestrated in a \gls{RT} scale, and how the near-\gls{RT} control loop interacts with the \emph{dApps}.

\section{Conclusions}\label{sec:Conclusions}
In this paper, we tackled the challenge of implementing efficient multi-time-scale control loops in \gls{O-RAN}-based deployments for \gls{uRLLC} services. We introduced \name{}, an \gls{O-RAN}-compliant framework specifically designed to address the radio resource allocation problem at both near-\gls{RT} and \gls{RT} scales. Central to our approach is the use of a novel Martingales-based model, to accurately compute the required number of guaranteed \glspl{RB} per service, ensuring that the violation probability---i.e., the probability that packet transmission delays exceed a predefined threshold---remains below the target tolerance. An additional key innovation in \name{} is the integration of an \gls{RT} control loop, which continuously monitors the transmission queues of each service and dynamically adjusts the allocation of guaranteed \glspl{RB} to address traffic anomalies in real time. 
Through extensive simulation, \name{} demonstrated significant improvements over existing solutions, achieving an average violation probability that is $\times 10$ lower than reference methods. This highlights the potential of our framework to enhance the reliability and efficiency of \gls{O-RAN}-based \gls{uRLLC} deployments, making it a promising approach for future networks.


\bibliographystyle{IEEEtran}
\bibliography{references}
\vskip -2\baselineskip plus -1fil
\begin{IEEEbiography}[{\includegraphics[width=1in,height=1.25in,clip,keepaspectratio]{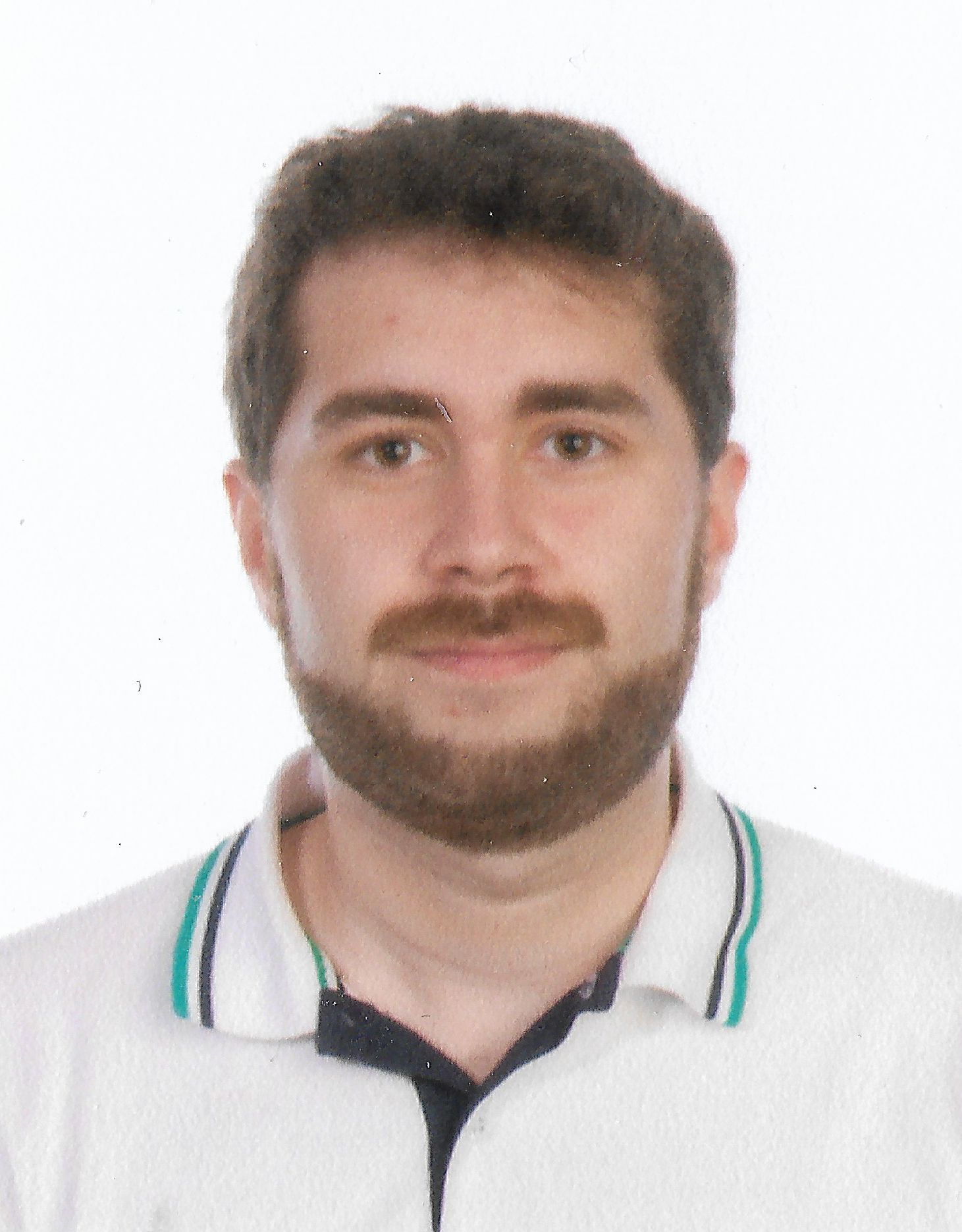}}]{Oscar Adamuz-Hinojosa} received the
B.Sc., M.Sc., and Ph.D. degrees in telecommunications engineering from the University of Granada, Granada, Spain, in 2015, 2017, and 2022, respectively. He was granted a Ph.D. fellowship by the Education Spanish Ministry in September 2018. He is currently an Interim Assistant Professor with the Department of Signal Theory, Telematics, and Communication, University of
Granada. He has also been a Visiting Researcher at NEC Laboratories Europe on several occasions. His research interests include network slicing, 6G radio access networks (RAN), and deterministic networks, with a focus on mathematical modeling.
\end{IEEEbiography}
\vskip -2\baselineskip plus -1fil 
\begin{IEEEbiography}[{\includegraphics[width=1in,height=1.25in,clip,keepaspectratio]{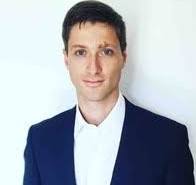}}]{Lanfranco Zanzi} received the B.Sc. and M.Sc. degrees in telecommunication engineering from the Polytechnic of Milan, Italy, in 2014 and 2017, respectively, and the Ph.D. degree from the Technical University of Kaiserlautern, Germany, in 2022. He works as a Senior Research Scientist with NEC Laboratories Europe. His research interests include network virtualization, machine learning, blockchain, and their applicability to 5G and 6G mobile networks.
\end{IEEEbiography}
\vskip -2\baselineskip plus -1fil
\begin{IEEEbiography}[{\includegraphics[width=1in,height=1.25in,clip,keepaspectratio]{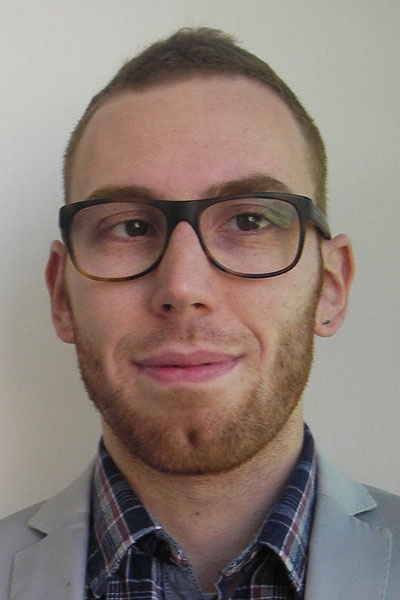}}]
{Vincenzo Sciancalepore} (M'15--SM'19) received his M.Sc. degree in Telecommunications Engineering and Telematics Engineering in 2011 and 2012, respectively, whereas in 2015, he received a double Ph.D. degree. Currently, he is a Principal Researcher at NEC Laboratories Europe, focusing his activity on reconfigurable intelligent surfaces. He is an Editor of the IEEE Transactions on Wireless Communications (since 2020) and IEEE Transactions on Communications (since 2024).
\end{IEEEbiography}
\vskip -2\baselineskip plus -1fil
\begin{IEEEbiography}[{\includegraphics[width=1in,height=1.25in,clip,keepaspectratio]{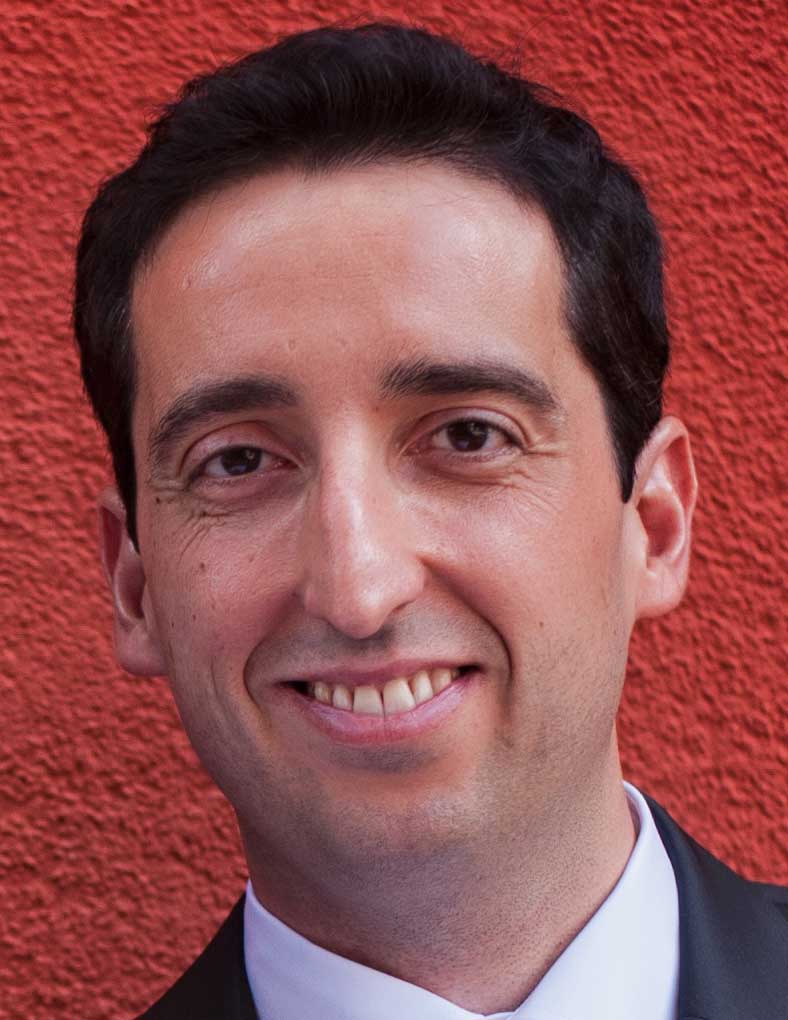}}]
{Xavier~Costa-P\'erez} (M'06--SM'18) is a Research Professor in ICREA, Scientific Director at the i2Cat Research Center and Head of 5G/6G Networks R\&D at NEC Laboratories Europe. He has served on the Organizing Committees of several conferences, published papers of high impact, and holds tenths of granted patents. Xavier received his  Ph.D. degree in Telecommunications from the Polytechnic University of Catalonia (UPC) in Barcelona and was the recipient of a national award for his Ph.D. thesis.
\end{IEEEbiography}

\end{document}